\documentclass[12pt, draftclsnofoot, onecolumn]{IEEEtran}

\IEEEoverridecommandlockouts
\usepackage{subfigure} 
\usepackage{graphicx}

\usepackage{amsmath,graphicx,amssymb,mathtools,bm}
\usepackage{subfigure}
\usepackage{hyperref}
\usepackage{cite}
\usepackage{amsmath,amssymb,amsfonts}
\usepackage{algorithmic}
\usepackage{graphicx}
\usepackage{textcomp}
\usepackage{xcolor}
\usepackage{verbatim}  
\usepackage{graphicx}  
\usepackage{bm}  
\usepackage{mathrsfs} 
\usepackage{algorithm} 
\usepackage{algorithmic} 
\usepackage{booktabs}
\usepackage{textcomp}  
\usepackage{multirow}  
\usepackage{lettrine}   
\usepackage{graphicx}  
\usepackage{color}  
\usepackage{amsmath}
\usepackage{amssymb}
\usepackage{stfloats}
\usepackage{caption}
\usepackage{subfigure}
\usepackage{pifont}
\usepackage{color}
\usepackage{setspace}

\begin{document}
	\newcommand{\tabincell}[2]{\begin{tabular}{@{}#1@{}}#2\end{tabular}}


\title{ Active 3D Double-RIS-Aided Multi-User Communications: Two-Timescale-Based Separate Channel Estimation via Bayesian Learning
}

\author{Songjie Yang, Wanting Lyu,
	Yue Xiu,  Zhongpei Zhang,~\IEEEmembership{Member,~IEEE}, and   Chau Yuen,~\IEEEmembership{Fellow,~IEEE}


\thanks{Songjie Yang, Wanting Lyu, Yue Xiu and Zhongpei Zhang are with the National Key Laboratory of Science and Technology on Communications, University of Electronic Science and Technology of China, Chengdu 611731, China. 
	(e-mail:
	yangsongjie@std.uestc.edu.cn;
	 lyuwanting@yeah.net;xiuyue12345678@163.com;
	zhangzp@uestc.edu.cn).

Chau Yuen is with the Engineering Product Development (EPD) Pillar, Singapore University of Technology and Design, Singapore 487372 (e-mail: yuenchau@sutd.edu.sg).}
}
\maketitle

\begin{abstract}
Double-reconfigurable intelligent surface (RIS) is a promising technique, achieving a substantial gain improvement compared to single-RIS techniques. However, in double-RIS-aided systems, accurate channel estimation is more challenging than in single-RIS-aided systems. This work solves the problem of double-RIS-based channel estimation based on active RIS architectures with only one radio frequency (RF) chain. Since the slow time-varying channels, i.e., the
 BS-RIS 1, BS-RIS 2, and RIS 1-RIS 2 channels, can be obtained with active RIS architectures, a novel multi-user two-timescale channel estimation protocol is proposed to minimize the pilot overhead. First, we propose an uplink training scheme for slow time-varying channel estimation, which can effectively address the double-reflection channel estimation problem. With channels' sparisty, a low-complexity Singular Value Decomposition Multiple Measurement Vector-Based Compressive Sensing (SVD-MMV-CS) framework with the line-of-sight (LoS)-aided off-grid MMV expectation maximization-based generalized approximate message passing (M-EM-GAMP) algorithm is proposed for channel parameter recovery.
For fast time-varying channel estimation, based on the estimated large-timescale channels,
a measurements-augmentation-estimate (MAE) framework is developed to decrease the pilot overhead.
Additionally, a comprehensive analysis of pilot overhead and computing complexity is conducted. Finally, the simulation results demonstrate the effectiveness of our proposed multi-user two-timescale estimation strategy and the low-complexity Bayesian CS framework.

\end{abstract}
\vspace{-1.15em} 
\section{Introduction}
\vspace{-0.15em} 
Reconfigurable intelligent surface (RIS) has been recognized as a potential future technology for sixth-generation (6G) communications \cite{6G1,6G2,6G3}. It can overcome obstacles, improve channel capacity, and lower transmit power for wireless communications.
With the fast advancement of the RIS technique in wireless communications, novel RIS structures have evolved, such as semi-passive/active RIS \cite{hybrid-ris1,hybrid-ris2,hybrid-ris3,hybrid-ris4,hybrid-ris6,hybrid-ris7}, double-/multi-RIS \cite{DR1,DR2,DR3,MR}, simultaneous transmitting and reflecting (STAR) RIS \cite{STAR1}, and holographic MIMO surfaces \cite{hol1,hol2}. To achieve the beamforming gain provided by RIS, accurate channel estimation is very essential.

As of now, a lot of researches have documented single-RIS-based channel estimation, with a focus on cascaded and separate channels. For cascaded channel estimation, the fully-passive RIS is considered. One straightforward approach is the ON/OFF switching method \cite{ONOFF}, for successive channel estimation of each RIS element. Following this, the full-ON methods, based on the discrete
Fourier transform (DFT) and Hadamard matrice, were proposed for better estimation \cite{FULLON}. Moreover, by exploiting the spatial sparsity of high-frequency channels, the pilot overhead can be reduced using compressive sensing (CS) methods. In \cite{CS1}, the authors transformed the cascaded channel estimation into a sparse recovery problem based on the Katri-Rao and/or Kronecker
products, and used the orthogonal match pursuit (OMP) algorithm to solve it. Furthermore, the cascaded channel's structural sparsity, i.e., row-column-block sparsity, was discussed in \cite{CS2}, for OMP-based multi-user channel estimation.  Along with this, more complex CS methods \cite{AMP1,AMP2, CS3}, e.g. message passing and atomic norm minimization, were used for RIS-based channel estimation.
  Another popular method is the matrix factorization/decomposition formulating cascaded channel estimation as a bilinear estimation problem \cite{PARAFAC1,PARAFAC2,PARAFAC3}. In this way, the cascaded channel estimation problem is decoupled into two separate channel estimation problems.
  Notably, the scaling ambiguity problem cannot escape. 
  That is, the matrix factorization/decomposition method is hard to acquire accurate separate channels. Moreover, deep learning methods have been applied for RIS-based channel estimation, such as deep neural network (DNN)- and convolutional neural network (CNN)-based cascaded channel estimation \cite{DL1,DL2}.

\begin{figure}
	\centering
	\includegraphics[width = 0.435\textwidth]{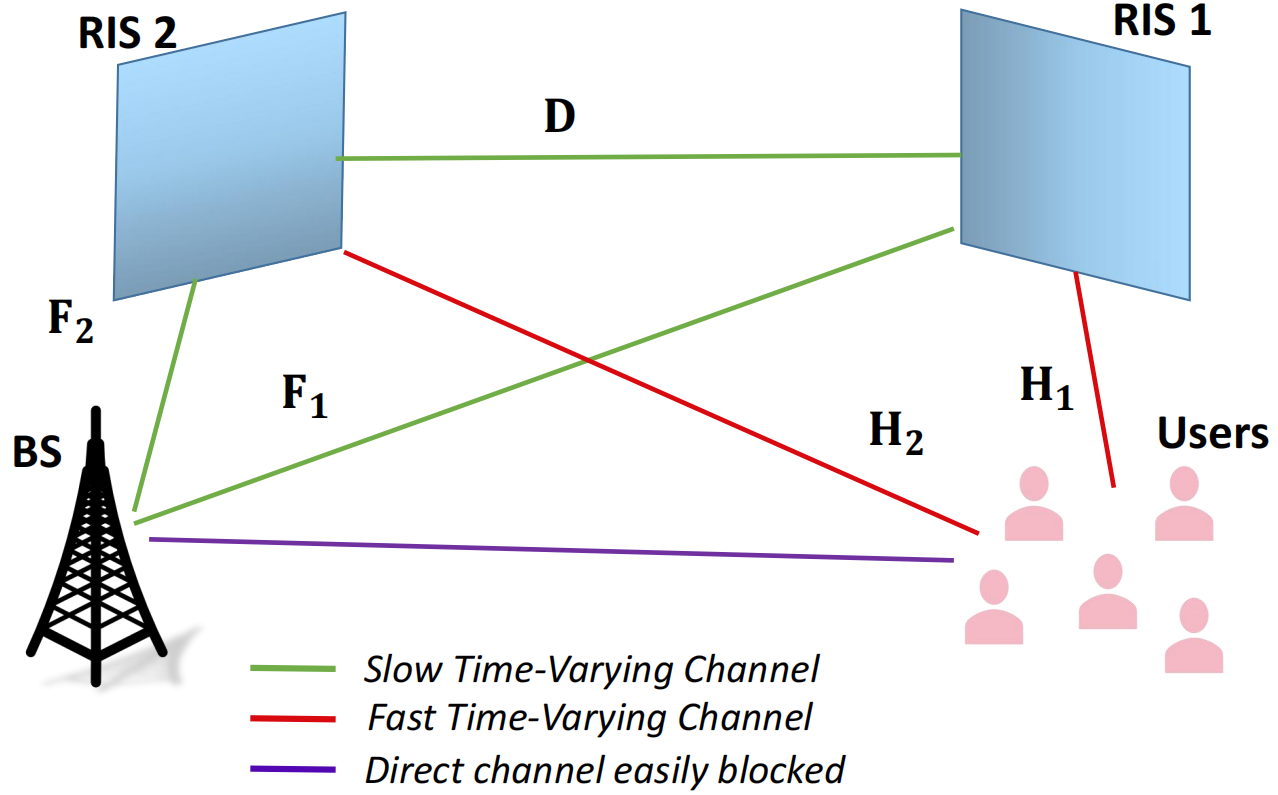}
	\caption{The architecture of double-RIS-aided systems.}
	\label{DRIS}
\end{figure}
On the other hand, the separate channel estimation issue must be handled to fulfill the requirements of situations requiring separate channels, such as \cite{separate2,separate3}. To this end, a few publications have investigated separate channel estimation with the semi-passive/active RIS architecture \cite{hybrid-ris1,hybrid-ris2,hybrid-ris3,hybrid-ris4,hybrid-ris6}. In \cite{hybrid-ris1}, a hybrid RIS with numerous active elements attached to radio frequency (RF) chains was designed to ease the channel estimation process. On this premise, the authors of \cite{hybrid-ris2} evaluated the estimation performance with varying numbers of RF chains, each connected to a single active element, using the CS approach. Due to the many RF chains, this design automatically results in a significant cost. In order to decrease hardware costs, \cite{hybrid-ris3,hybrid-ris4} used an architecture that connected a single RF chain to numerous active elements. Notably, all these strategies employed a two-way method. Specifically, both the base station (BS) and the user transmit pilots to the RIS in order to estimate the channels. The RIS then transmits the channel information to the BS over the wired control. This technique focuses on the active RIS's capacity to receive signals, but ignores the fact that it can also reflect signals, resulting in additional pilot overhead. To solve this, the authors of \cite{hybrid-ris6,hybrid-ris7} presented a one-way training strategy that reduced training overhead by taking into account the RIS's capacity to reflect.

Recently, the double-RIS design was widely studied in \cite{DR1,DR2,DR3}, with one RIS set near the BS and the other RIS set near the user, for capacity enhancement. 
This architecture could achieve an $L^4$-fold cooperative power gain (a substantial improvement compared with the single-RIS aided system's $L^2$-fold gain) by the double-reflection link \cite{DR1}, with $L$ representing the number of RIS elements. 
Notably, in such a double-RIS-aided system, channel estimation will become more problematic due to two single-reflection channels and one double-reflection channel. In \cite{DR1}, two semi-passive RISs were assumed to receive signals for estimating the channels between them and the BS/user. To deal with the double-reflection channel, the assumption is made that the inter-RIS channel consisted only of the known LoS path. This may be impractical in outdoor communications. To tackle the double-reflection channel,
 the authors in \cite{DRCE1,DRCE2,DRCE3,DRCE4,DRCE5} have presented effective cascaded channel estimation methods for fully-passive double-RIS systems.
 Nevertheless, these methods will result in a high amount of training complexity, and obtaining separate channels will be challenging. Earlier-mentioned semi-passive/active RIS can make up for this by incurring low hardware expenses. In this sense, the constraint of \cite{DR1} with semi-passive RISs should be overcome. Additionally, as in the case of single-RIS-based channel estimation, the sparsity of the high-frequency channel could be utilized to reduce the pilot overhead.

Last but not least, the assumption that both the BS and the RIS are equipped with uniform planar arrays (UPAs) is more realistic than the case of uniform linear arrays considered in the majority of CS-based channel estimation \cite{CS1,CS2,AMP1,AMP2,hybrid-ris2,hybrid-ris6}. This UPA-based system setup \cite{hybrid-ris3,hybrid-ris4}, however, will make it more challenging to estimate the channel's parameters due to the three-dimensional (3D) beampattern, especially in double-RIS-aided systems. In this case, the conventional CS framework, i.e. Kronecker CS, is relatively time-consuming when used to the multi-parameter recovery issue due to the immense dictionary generated by the vectorization of multidimensional signals. Such a recovery framework, with a large computational cost, is impracticable for UPA-based systems when recovering the 3D beamspace (the BS's and RIS's angles-of-depature (AoDs)/angles-of-arrival (AoAs) in the elevation and azimuth directions). Therefore, it is essential to investigate estimation methods with minimal complexity for 3D double-RIS-aided systems.

Taking into account the above, an efficient separate channel estimation strategy for active 3D double-RIS-aided systems is investigated in this paper, in which the BS and the RISs are all equipped with UPAs.
Compared with  \cite{DRCE1,DRCE2,DRCE3,DRCE4,DRCE5}, this work can obtain separate channels owing to the RF chain set at the RIS. Besides, the channel's sparsity is exploited to effectively reconstruct the channel.
 Based on the separate channels, the two-timescale estimation technique, which has been employed in relay- and single-RIS-aided systems \cite{twoscale1,twoscale2}, is used for pilot minimization.
 In comparison to the most related work \cite{DR1} with semi-passive RISs, the double-reflection channel estimation problem is addressed.

 Before presenting our work, we first introduce three significant channel properties of double-RIS-aided systems.
\begin{itemize}
	\item \textbf{Two-Timescale Channel Property:}  Owing to the fixed locations of the BS and RISs, the channels among them are slow time-varying. On the contrary, the channel between the RIS and the mobile user is fast time-varying. 
	\item \textbf{LoS-Path Property:} Owing to the known locations of the BS and RISs, the LoS paths' AoA/AoD of channels among them can be easily obtained.
	\item \textbf{Multi-User Channel Property:} For each RIS, all users share the same BS-RIS channel. This indicates that the BS's AoD and the RIS's AoA are common for all users' channels.
\end{itemize}
According to the above properties, the following contributions are exhibited with respect to hardware cost, pilot overhead and computational complexity:

\begin{itemize}
	\item \emph{Cost-Minimized Active RIS Structure:} Considering the hardware cost of the active RIS, we use only one RF chain to receive signals. Different from the active RIS structure with each active element connected to a RF chain \cite{hybrid-ris1,hybrid-ris2,hybrid-ris6}, the single-RF-chain-based RIS makes each RF chain connects to all elements. Since the proposed channel estimation method makes no power requirements of the RIS, all active elements can be set to unit power, which is similar to the fully-passive RIS, without incurring additional power consumption.

\item \emph{Pilot-Minimized Channel Estimation Protocol:} Considering that the single-RF-chain is bound to incur higher training overhead for parameter recovery compared with multiple RF chains, a novel multi-user two-timescale estimation technique is proposed to minimize the pilot overhead based on the CS theory\footnote{When using the CS method for high-frequency channels, the channel estimation problem can be converted into a beamspace recovery problem.} and \textbf{Two-Timescale Channel Property}. In this sense, we first estimate the slow time-varying channels in an uplink training manner, which can effectively solve the double-reflection channel estimation problem in \cite{DR1}. Leveraging on \textbf{Multi-User Channel Property}, a   measurements-augmentation-estimate (MAE) CS strategy\footnote{In a CS framework, it may be recognized that the observed values rise for some underlying reasons.} 
is proposed for better estimation.
Based on the estimated slow time-varying channels, the fast time-varying channels can be estimated at the BS with more RF chains, so as to develop a MAE-based solution with low pilot overhead.
\item \emph{Computational Complexity-Minimized CS Framework:}
 To avoid the unacceptable computational complexity induced by recovering the 3D beamspace, we
proposed a novel CS framework for multi-parameter recovery. Specifically,
inspired by the decoupling of two parameter recovery problems using SVD in \cite{SVD}, we exploit the common sparsity of decomposition vectors to develop a more effective singular value decomposition multiple measurement vector-based compressive sensing (SVD-MMV-CS) Framework. Most importantly, it is much faster and has better performance than the common Kronecker CS framework, as shown in simulation results.
Besides, the LoS-aided dictionary based on \textbf{LoS-Path Property} is designed, which can resist the basis mismatch problem \cite{mismatch} to some extent.
\item \emph{Efficient Bayesian Recovery Algorithm:} With our proposed CS framework, the selection of the recovery algorithm is very crucial. 	To achieve an excellent recovery performance, the Bayesian learning-based method is considered for the 3D beamspace recovery.
Given the great computing cost produced by the inverse operation of the standard Bayesian recovery method, i.e., EM-based sparse Bayesian learning, the expectation maximization-based generalized approximate message passing (EM-GAMP) \cite{EM-GAMP} algorithm is used, which utilizes the GAMP procedure to approximate the true posterior distribution with low complexity.
To employ the EM-GAMP algorithm for our proposed CS framework, we extend it to the MMV version (M-EM-GAMP) by sharing a sparse rate factor across all measurement vectors.  
Considering that M-EM-GAMP cannot handle the mismatch problem, we propose a fast off-grid MMV method, based on the Taylor expansion, to refine the on-grid recovered parameters. 
\end{itemize}

The rest of this paper is organized as follows: in Section \ref{S2}, we model the two-timescale channels and describe the received signals at the BS and RISs via uplink training.
Section \ref{S3} shows the two-timescale channel estimation protocol, and design the CS recovery problems for the large-timescale and small-timescale estimation, respectively. Section \ref{S4} proposes a LoS-aided dictionary, a low-complexity CS framework, an M-GAMP algorithm, an EM procedure for updating the hyperparameters of M-GAMP, and the Taylor expansion-based off-grid refinement method.  Section \ref{S5} evaluates the large-timescale and small-timescale channel estimation methods via various experiments. Additionally, the analysis of pilot overhead and computational complexity of different schemes is provided. Finally, Section \ref{S6} concludes the paper.

{\emph {Notations}}:
 ${\left(  \cdot  \right)}^{ *}$, ${\left(  \cdot  \right)}^{ T}$ and ${\left(  \cdot  \right)}^{ H}$ denote conjugate, transpose, conjugate transpose, respectively. ${{\mathbf{A}}^\dag }$ is the Moore-Penrose pseudoinverse matrix. $\Vert\cdot\Vert_0$ and $\Vert\cdot\Vert_2$ represent $\ell_0$ norm and $\ell_2$ norm, respectively. 
 $\Vert\mathbf{A}\Vert_F$ denotes the Frobenius norm of matrix $\mathbf{A}$. Furthermore, $\otimes$ and $\odot$ are the Kronecker product and the Hadamard product, respectively. $[\mathbf{a}]_{i}$ and $[\mathbf{A}]_{i,j}$ denote the $i$-th element of vector $\mathbf{a}$, the $(i,j)$-th element of matrix $\mathbf{A}$, respectively. $\rm{vec}$ represents the vectorization operation.
 $\textbf{Re}\{a\}$ denotes the real part of complex $a$. $\mathbb{E}\{\cdot
 \}$ and $\mathbb{V}\{\cdot\}$ denote the expectation and variance operations , respectively.  $\mathbf{I}_M$ denotes the $M$-by-$M$ identity matrix. Moreover, $\rm{diag}(\mathbf{a})$ is a square diagonal matrix with entries of $\mathbf{a}$ on its diagonal. Finally, $\mathcal{CN}(\mathbf{a},\mathbf{A})$ is the complex Gaussian distribution with mean $\mathbf{a}$ and covariance matrix $\mathbf{A}$.

%

\section{System and Channel Model}
\label{S2}
\begin{figure}
	\begin{minipage}[t]{0.51\linewidth}
		\centering
		\subfigure[Fully-Connected Structure]{\includegraphics[height=2.85cm,width=6cm]{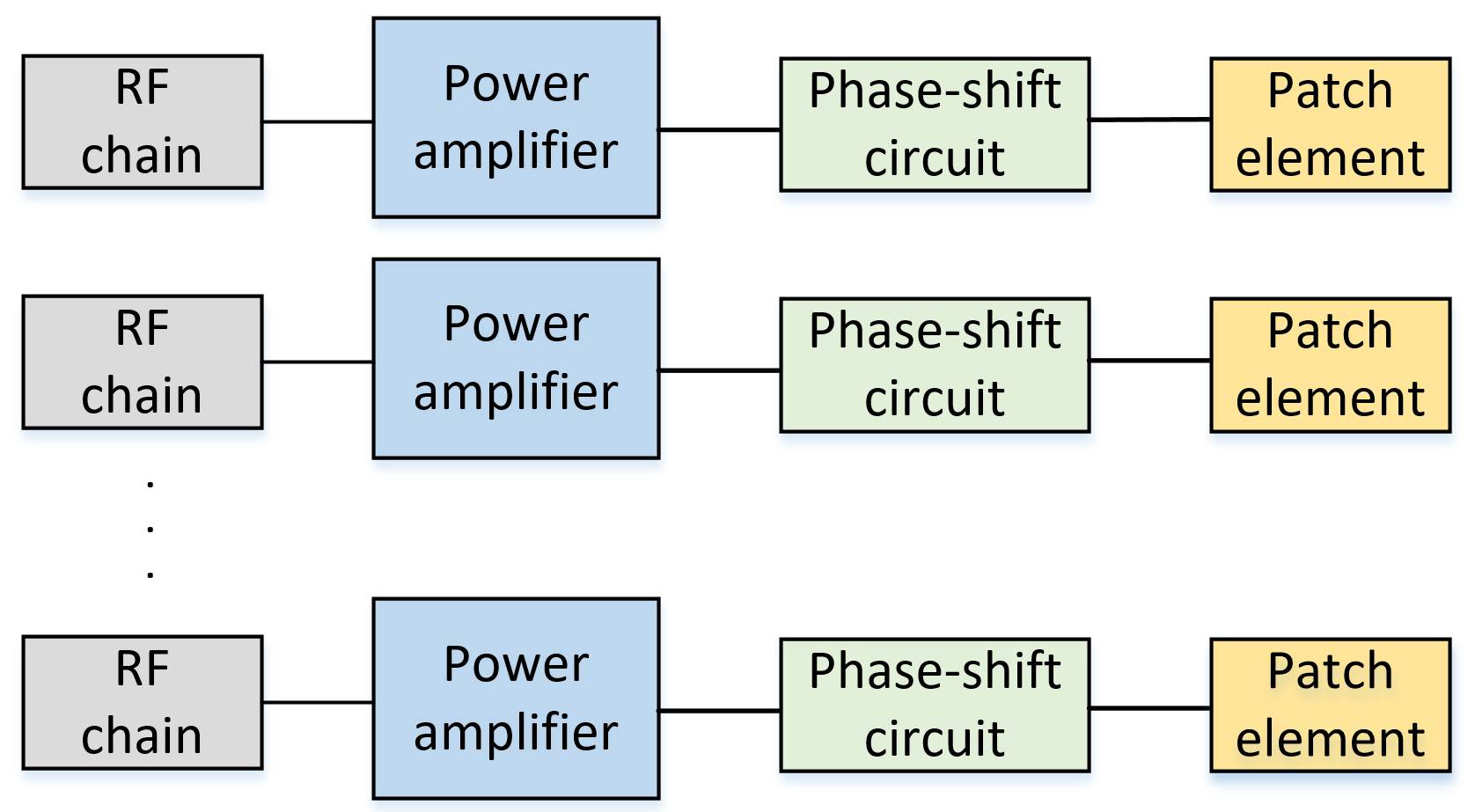}}
	\end{minipage}%
	\begin{minipage}[t]{0.51\linewidth}
		\centering
		\subfigure[Sub-Connected Structure]{\includegraphics[height=2.75cm,width=6.3cm]{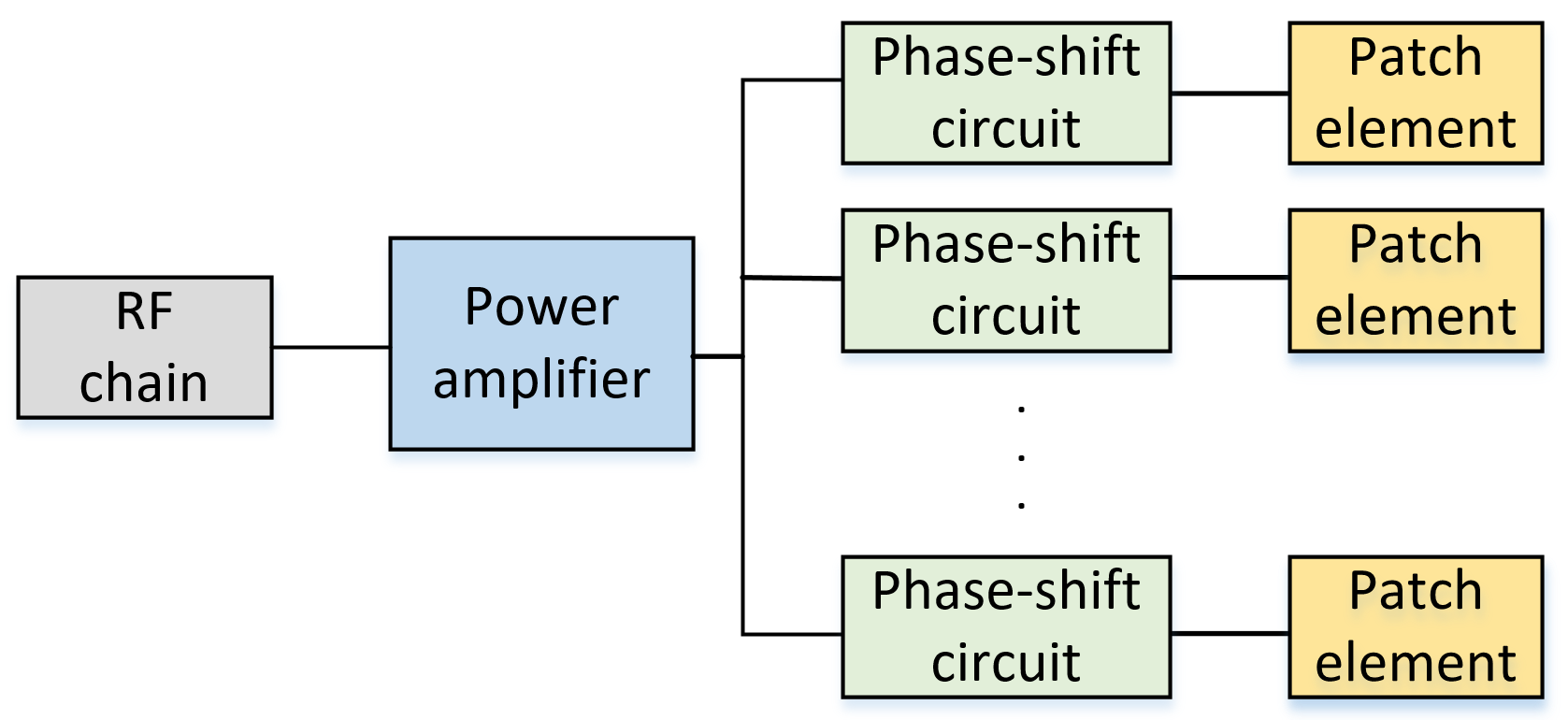}}
	\end{minipage}
	\caption{Two active RIS Structures with the RF chain.}
	\label{active}
\end{figure}

We consider a multi-user double-RIS-aided system with central frequency $28 \ {\rm GHz}$, shown in Fig. \ref{DRIS}, where the BS with the assistance of the RIS serves $U$ users. The BS is equipped with $J$ antenna elements, and each of RISs is equipped with $L$ reflective elements. The users are all equipped with a single antenna. As shown in Fig. \ref{DRIS}, we denote the channels from the BS to RIS $i$ as $\mathbf{F}_i$, from RIS $1$ to RIS $2$ as $\mathbf{D}$, and from the RIS $i$ to the $u$-th user as $\mathbf{h}_{i,u}\in\mathbb{C}^{L\times 1}$, where $i=1,2$ and $u=1,\cdots,U$. 
 Since there is no requirement that the power of each element is controlled separately for channel estimation, the situation shown in Fig. \ref{active}(a) is unnecessary.
 Consequently, we utilize a unique low-cost active RIS structure with one RF chain and one power amplifier (PA), as shown in Fig. \ref{active}(b). This is a specific instance of the active RIS with the sub-connected architecture introduced in \cite{hybrid-ris5}. Next, Section. \ref{Channel Model} models the large-timescale and small-timescale channels, and Section. \ref{Signal Model} gives the received signal model via uplink training.
\subsection{Two-Timescale Channel Model}\label{Channel Model}

 Quasi-static block-fading channels are considered, we adopt the widely used spatial channel model, characterizing the geometrical structure.
Assuming that the BS and RISs are all equipped with UPAs, then the slow and fast time-varying channels are modeled as follows.
\subsubsection{Slow Time-Varying Channel} The slow time-varying channels between the BS and RIS 1, the BS and RIS 2, and RIS 1 and RIS 2, i.e., $\mathbf{F}_1$, $\mathbf{F}_2$ and $\mathbf{D}$, are respectively expressed as
\begin{equation}
	\setlength{\abovedisplayskip}{2.9pt}
\begin{aligned}
		{\mathbf{F}}_i\!= \!\sqrt {\frac{{L\!J}}{{P_{f_i}}}} \sum\limits_{p = 1}^{{P_{f_i}}}\! \ {\alpha_{p,i}}  {{\mathbf{a}}_{\rm B}}\left(  \frac{2d}{\lambda}{\cos(\vartheta^{\rm ele}_{p,i})},\frac{2d}{\lambda}\sin(\vartheta^{\rm ele}_{p,i})\sin(\vartheta^{\rm azi}_{p,i})  \right) 
		  {{\mathbf{a}}_{\rm L_i}^{H}}\left(  \frac{2d}{\lambda}{\cos(\varphi ^{\rm ele}_{p,i}),\frac{2d}{\lambda}\sin(\varphi_{p,i}^{\rm ele})\sin(\varphi_{p,i}^{\rm azi})}  \right),
\end{aligned}
		\setlength{\belowdisplayskip}{2.9pt}
\end{equation}
\begin{equation}\label{D}
	\setlength{\abovedisplayskip}{2.9pt}
	\begin{aligned}
	{\mathbf{D}}\!= \!\sqrt {\frac{{L^2}}{{P_{d}}}} \sum\limits_{b = 1}^{{P_{d}}}\!  \ \beta_b
	{{\mathbf{a}}_{\rm L_2}}\left(  \frac{2d}{\lambda}{\cos(\theta ^{\rm ele}_{b}),\frac{2d}{\lambda}\sin(\theta_b^{\rm ele})\sin(\theta_b^{\rm azi})}  \right)   {{\mathbf{a}}_{\rm L_1}^{H}}\left(  \frac{2d}{\lambda}{\cos(\phi ^{\rm ele}_{b}),\frac{2d}{\lambda}\sin(\phi_b^{\rm ele})\sin(\phi_b^{\rm azi})}  \right),
\end{aligned}
		 \setlength{\belowdisplayskip}{2.9pt}
\end{equation}
where $i\in\{1,2\}$ denote the indices of the two RISs. $P_{f_i}$ and $P_d$ are the number of paths of $\mathbf{F}_i$ and $\mathbf{D}$, respectively. $\{\vartheta^{\rm ele/azi}_{p,i}\}_{p=1}^{P_{f_i}}$ and $\{\varphi^{\rm ele/azi}_{p,i}\}_{p=1}^{P_{f_i}}$ represent AoDs and AoAs of the channel between the BS and RIS $i$, respectively. Notably, the superscript $\rm ele$ and $\rm azi$ denote the elevation and azimuth directions, respectively.
$\{\theta_b^{\rm ele/azi}\}_{b=1}^{P_d}$ and $\{\phi_b^{\rm ele/azi}\}_{b=1}^{P_d}$ are AoDs and AoAs of the channel between RIS 2 and 1, respectively.
 $\{\alpha_{p,i}\}_{p=1}^{P_{f_i}}$ and $\{\beta_b\}_{b=1}^{P_d}$ are the path gains of the above channels. 
In addition, $\lambda$ is the antenna wavelength, $d$ is the antenna inter-element spacing, generally $d=\lambda/2$. Ignoring the subscript, the UPA steering  vector $\mathbf{a}(x_1,x_2)$ is given by
\begin{equation}\label{UPAs}
	\setlength{\abovedisplayskip}{3.2pt}
	\mathbf{a}(x_1,x_2)=\sqrt{\frac{1}{N_yN_z}}\left[1,e^{j\pi x_1},\cdots,e^{j\pi(N_z-1)x_1}\right]^T\otimes\left[1,e^{j\pi x_2},\cdots,e^{j\pi(N_y-1)x_2}\right]^T,
	\setlength{\belowdisplayskip}{3.2pt}
\end{equation}
where $N_y$ and $N_z$ are the number of antennas in the $y$ and $z$ axis, respectively.
\subsubsection{Fast Time-Varying Channel}
Due to the mobility of users, the channels with respect to (w.r.t.) users are thought of fast time-varying. The channels between the BS and users, and between RISs and users are 
$
{{\mathbf{h}}_{i,u}}= 
\sum\limits_{c = 1}^{{P_{h,i,u}}} {{\bar{\gamma}_{i,u,c}}{\mathbf{a}}_{\rm L_i}\left( \frac{2d}{\lambda} \cos({\chi_{i,u,c}^{\rm ele}}),\frac{2d}{\lambda} \sin({\chi_{i,u,c}^{\rm ele}})
	\sin({\chi_{i,u,c}^{\rm azi}}) \right)}
$,
where ${\bar{\gamma}_{i,u,c}}=\sqrt {\frac{{L}}{{P_{h,i,u}} }}{\gamma_{i,u,c}}$, 
 $P_{h,i,u}$ is the number of paths of $\mathbf{h}_{i,u}$, $\{\gamma_{i,u,c}\}_{c=1}^{P_{h,i,u}}$ and $\{\chi_{i,u,c}^{\rm ele/azi}\}_{c=1}^{P_{h,i,u}}$ denote path gains and AoDs from RIS $i$ to user $u$, respectively.

\subsubsection{Virtual Channel Representation}
Using the virtual channel representation, the BS-RIS, RIS 1-2 and RISs-user channels can be rewritten as
\begin{equation}\label{VCR_h}
	\setlength{\abovedisplayskip}{3.2pt}
	\mathbf{F}_i\approx\mathbf{A}_{\rm B}\bm{\Xi}_i\mathbf{A}_{\rm L_i}^H, \
	\mathbf{D}\approx\mathbf{A}_{\rm L_2}\bm{\Delta}\mathbf{A}_{\rm L_1}^H, \
	\mathbf{h}_{i,u}\approx\mathbf{A}_{\rm L_i}\bm{\zeta}_{i,u},
	\setlength{\belowdisplayskip}{3.2pt}
\end{equation}
where $\mathbf{A}_{\rm  B}\in\mathbb{C}^{J\times G_t}$ and $\mathbf{A}_{\rm L_i}\in\mathbb{C}^{L\times G_r}$ denote the dictionaries of the BS's AoD, RISs' AoA/AoD, respectively.  $\bm{\Xi}_i\in\mathbb{C}^{G_t\times G_r}$ denotes the sparse matrix, where the non-zero entries correspond to the BS's AoDs and the $i$-th RIS's AoAs. This resembles both $\mathbf{D}$ and $\mathbf{h}_{i,u}$. The later section will provide specific dictionaries appropriate for RIS systems.

\vspace{-0.75em}
\subsection{Uplink Training}\label{Signal Model}
We arrange the uplink training into two parts based on the estimates of $\{\mathbf{F}_i,\mathbf{D}\}_{i=1}^2$ and $\{\{\mathbf{h}_{i,u}\}_{i=1}^2\}_{u=1}^U$, respectively. 
For RIS-based multi-user channel estimation, the key is to design orthogonal pilot sequences and reflection coefficients. Assume that there are $Q$ sub-frames for uplink training and $T$ symbol durations for users in each sub-frame, where $T\geq U$. Owing to  orthogonal pilot sequences, we have $\mathbf{s}^H_{u_1}\mathbf{s}_{u_2}=0, u_1\neq u_2$ and $\mathbf{s}^H_{u_i}\mathbf{s}_{u_i}=\sigma_p^2T, \forall
 i$, where $\sigma_p^2$ is the transmit power. Denoting the reflection coefficients of RIS 1 and RIS 2 in the $q$-th sub-frame and the pilot sequence of the $u$-th user by $\mathbf{v}_{1,q}\in\mathbb{C}^{L\times 1}$, $\mathbf{v}_{2,q}\in\mathbb{C}^{L\times 1}$ and $\mathbf{s}_u^H\in\mathbb{C}^{1\times T}$, respectively. 
In the $q$-th sub-frame, the received signals at RIS 1, RIS 2 and the BS are denoted by $\mathbf{y}_q^{\rm RIS_1}$, $\mathbf{y}_q^{\rm RIS_2}$ and $\mathbf{Y}_q^{\rm BS}$. They are given by
 \begin{equation}
 	\setlength{\abovedisplayskip}{3.2pt}
	\mathbf{y}_q^{\rm RIS_1}=\sum\nolimits_{u=1}^{U}\mathbf{v}_{1,q}^{H}\mathbf{h}_{1,u}\mathbf{s}_u^H+\mathbf{v}_{1,q}^{H}\mathbf{N}^{{\rm RIS}_1}_q,
	\setlength{\belowdisplayskip}{3.2pt}
\end{equation}
 \begin{equation} 		\setlength{\abovedisplayskip}{3.2pt}
	\mathbf{y}_q^{\rm RIS_2}=\sum\nolimits_{u=1}^{U}\mathbf{v}_{2,q}^{H}\mathbf{D}\mathbf{V}_{1,q}\mathbf{h}_{1,u}\mathbf{s}_u^H+\mathbf{v}^{H}_{2,q}\mathbf{h}_{2,u}\mathbf{s}_u^H
	+\mathbf{v}_{2,q}^{H}\mathbf{N}^{{\rm RIS}_2}_{q},
	\setlength{\belowdisplayskip}{3.2pt}
\end{equation}
 \begin{equation}
 	\setlength{\abovedisplayskip}{3.2pt}
 	\mathbf{Y}_q^{\rm BS}=\sum\nolimits_{u=1}^{U}
 	(\mathbf{F}_1\mathbf{V}_{1,q}\mathbf{h}_{1,u}+\mathbf{F}_2\mathbf{V}_{2,q}\mathbf{h}_{2,u}
 	+\mathbf{F}_2\mathbf{V}_{2,q}\mathbf{D}\mathbf{V}_{1,q}\mathbf{h}_{1,u})\mathbf{s}_u^H
  +\mathbf{N}_{q}^{\rm BS},
  	\setlength{\belowdisplayskip}{3.2pt}
\end{equation}
where $\mathbf{V}_{i,q}={\rm diag}(\mathbf{v}_{i,q})\in\mathbb{C}^{L\times L}, \ i\in\{1,2\}$. $\mathbf{N}_{q}^{{\rm RIS}_1}\in\mathbb{C}^{L\times T}$, $\mathbf{N}_{q}^{{\rm RIS}_2}\in\mathbb{C}^{L\times T}$ and $\mathbf{N}_q^{\rm BS}\in\mathbb{C}^{J\times T}$ are the Gaussian white noise.

In contrast to the received signal from the RIS, the signal delivered to the BS is comprised of the RIS's reflected signal and the direct signal. Since the direct channel between the BS and the user can be estimated simply by turning off all RISs, the direct signal can be removed from the BS's received signal. As a result, this study need not consider the direct channel. This, too, is built on the same considerations as cascaded channel estimation literature.

\begin{figure*}
	\centering
	\includegraphics[width = 0.9\textwidth]{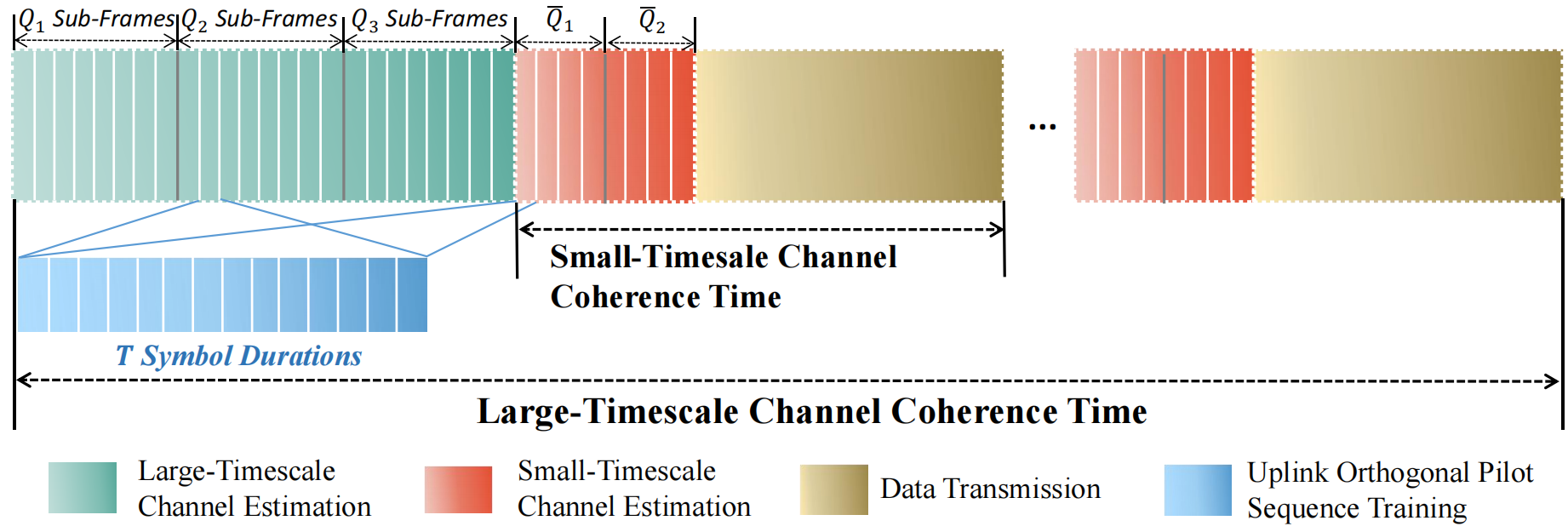}
	\caption{Multi-user two-timescale channel estimation protocol for double-RIS-aided systems.}
	\label{CEprotocol}
\end{figure*}
\section{Two-Timescale-Based Separate Channel Estimation Framework}\label{S3}
In this section, we propose a two-timescale-based separate channel estimation framework using \textbf{Two-Timescale Channel Property} and \textbf{Multi-user Channel Property}. For large-timescale estimation, a three-phase scheme is proposed to estimate $\mathbf{F}_1$, $\mathbf{F}_2$ and $\mathbf{D}$ separately. Different from the approach in \cite{hybrid-ris1,hybrid-ris2,hybrid-ris3,hybrid-ris4} that the BS transmits the pilot to estimate the BS-RIS channel at the RIS, our proposed method exploits uplink training sequences sent from users to estimate $\mathbf{F}_1$, $\mathbf{F}_2$ and $\mathbf{D}$ at the BS. There are two benefits of this approach. First, the measurement dimension of the signal received by the BS is greater than that received by the RIS due to the number of the BS's RF chains are much larger than that of RISs'. The second is joint multi-user channel estimation. Since all users share the common channels $\{\mathbf{F}_1,\mathbf{F}_2,\mathbf{D}\}$, as shown in Fig. \ref{DRIS}, the BS's received signals from all users can be treated as observations of the same space. This indicates that multi-user signals can be processed jointly to improve recovery performance. In this way, a MAE strategy based on \textbf{Multi-User Channel Property} is proposed. Specifically, since the channel $\mathbf{F}_i$ is common for all users, the signals of different users received by the BS can be utilized to jointly estimate their shared channel $\mathbf{F}_i$.
 For small-timescale estimation, we also present a MAE strategy. Based on the channels estimated in the large-timescale,  $\{\{\mathbf{h}_{i,u}\}_{i=1}^2\}_{u=1}^U$ can be estimated at the BS with numerous RF chains as opposed to the RIS, in order to improve recovery performance.
 
 We assume that there are $Q_1$, $Q_2$, $Q_3$, $\bar{Q}_1$ and $\bar{Q}_2$ sub-frames for the estimation of $\mathbf{F}_1$, $\mathbf{F}_2$, $\mathbf{D}$, $\{\mathbf{h}_{1,u}\}_{u=1}^U$ and
 $\{\mathbf{h}_{2,u}\}_{u=1}^U$, respectively. Similar to the previous description, each sub-frame consists of $T \ (T\geq U)$ sysmbol durations for orthogonal pilot transmission. Following that, the proposed multi-user two-timescale channel estimation protocol, accommodating large-timescale channel estimation, small-timescale channel estimation, data transmission, and uplink orthogonal pilot sequence training,
  is shown in Fig. \ref{CEprotocol}.
  It can be noticed that 
 in a large-timescale period, the slow time-varying channels just need to be estimated only once. 
  Therefore, the whole pilot overhead is reduced, compared with the traditional estimation protocol that needs to estimate the slow and fast time-varying channels with the same number of times. 
  Next, we will introduce the large-scale and small-scale estimation methods in detail. 

\subsection{Large-Timescale Estimation}\label{largescale}
\subsubsection{Estimation of $\mathbf{F}_i$}
For simplicity, here we use the time-division mode, in which one RIS is active, while another is inactive. Utilizing the receiving/processing signal capacity of the RIS to acquire the RIS-user channel and subsequently the reflection capability of the RIS to estimate $\mathbf{F}_i$ at the BS using the estimated RIS-user channel is essential to our suggested MAE strategy.

First, the $i$-th RIS's received signal 
$\mathbf{y}_{q_i}^{{\rm RIS}_i}\in\mathbb{C}^{1\times T}$ in the $q_i$-th sub-frame, with $q_i\in\{1,\cdots,Q_i\}$, is given by
$
	\mathbf{y}_{q_i}^{{\rm RIS}_i}=\sum\nolimits_{u=1}^{U}\mathbf{v}_{i,q_i}^H\mathbf{h}_{i,u}\mathbf{s}_u^H+\mathbf{v}_{i,q_i}^H\mathbf{N}^{{\rm RIS}_i}_{q_i}
$,
where $\mathbf{v}_{i,q_i}$ is the $i$-th RIS's reflect coefficients in the $q_i$-th sub-frame, and $\mathbf{N}^{{\rm RIS}_i}_{q_i}\in\mathbf{C}^{L\times T}$ is the Gaussian noise vector following $\mathcal{CN}(0,\sigma_n^2\mathbf{I}_T)$.
The collected received signal $\widetilde{\mathbf{Y}}^{{\rm RIS}_i}\in\mathbb{C}^{Q_i\times T}$ in $Q_i$ sub-frames at RIS $i$ can be expressed as
\begin{equation}\label{RIS_Y}
	\setlength{\abovedisplayskip}{3.2pt}
	\widetilde{\mathbf{Y}}^{{\rm RIS}_i}=\sum\nolimits_{u=1}^{U}\mathring{\mathbf{V}}_i^H\mathbf{h}_{i,u}\mathbf{s}_u^H+\widetilde{\mathbf{N}}^{{\rm RIS}_i},
	\setlength{\belowdisplayskip}{3.2pt}
\end{equation}
where $\mathring{\mathbf{V}}_i=[\mathbf{v}_{i,1},\cdots,\mathbf{v}_{i,Q_i}]\in\mathbb{C}^{L\times Q_i}$ and $\widetilde{\mathbf{N}}^{{\rm RIS}_i}=[\mathbf{v}_{i,1}^H\mathbf{N}^{{\rm RIS}_i}_{1},\cdots,\mathbf{v}_{i,Q_i}^H\mathbf{N}^{{\rm RIS}_i}_{Q_i}]^T\in\mathbb{C}^{Q_i\times T}$.
Owing to the orthogonal pilot sequences, we obtain the $u$-th user's received signal $\widetilde{\mathbf{y}}_u^{{\rm RIS}_i}\in\mathbb{C}^{Q_i\times 1}$, i.e.,
\begin{equation}\label{RIS_yu}
	\setlength{\abovedisplayskip}{3.2pt}
	\widetilde{\mathbf{y}}_u^{{\rm RIS}_i}=\frac{1}{\sigma_p^2T}\widetilde{\mathbf{Y}}^{{\rm RIS}_i}\mathbf{s}_u 
=\mathring{\mathbf{V}}_i^H\mathbf{h}_{i,u}+\widetilde{\mathbf{n}}^{{\rm RIS}_i}_{u}\overset{(a)}{\approx}\mathring{\mathbf{V}}_i^H\mathbf{A}_{\rm L_i}\bm{\zeta}_{i,u}+\widetilde{\mathbf{n}}^{{\rm RIS}_i}_{u}
	\setlength{\belowdisplayskip}{3.2pt}
\end{equation}
where $(a)$ holds because of Eqn. (\ref{VCR_h}),  $\widetilde{\mathbf{n}}^{{\rm RIS}_i}_{u}=\frac{1}{\sigma_p^2T}\widetilde{\mathbf{N}}^{{\rm RIS}_i}\mathbf{s}_u\in\mathbb{C}^{Q_i\times 1}$ and $\mathring{\mathbf{V}}_i^H\mathbf{A}_{\rm L_i}\in\mathbb{C}^{Q_i\times G_r}$ is the sensing matrix.
 $\{\mathbf{h}_{i,u}\}_{u=1}^U$ can be easily estimated by a variety of recovery algorithms. Then they are sent to the BS as the auxiliary information for estimating $\mathbf{F}_i$ through the wired control link\footnote{Also, there are two additional types of feedback: pilot feedback used to estimate $\{\widehat{\mathbf{h}}_u\}_{u=1}^U$ at the BS and estimated spatial parameter (angle and gain information) feedback used to rebuild the channel at the BS.}.

Similar to the previous process,
 the BS's received signal reflected by RIS $i$ in the $q_i$-the sub-frame is given by
 \begin{equation}\label{YBSi}
 	\setlength{\abovedisplayskip}{3.2pt}
 	\widetilde{\mathbf{y}}^{\rm BS}_{i,u,q_i}=\mathbf{F}_i\mathbf{V}_{i,q_i}\mathbf{h}_{i,u}+\widetilde{\mathbf{n}}_{i,u,q_i}^{\rm BS}
 	=\mathbf{F}_i{\rm diag}(\mathbf{h}_{i,u})\mathbf{v}_{i,q_i}+\widetilde{\mathbf{n}}_{i,u,q_i}^{\rm BS},
 	\setlength{\belowdisplayskip}{3.2pt}
 \end{equation}
where $\widetilde{\mathbf{n}}_{i,u,q_i}^{\rm BS}$ is noise.
 Hence, the BS's received signal in $Q_i$ sub-frames is expressed as
\begin{equation}\label{YIU}
	\setlength{\abovedisplayskip}{3.2pt}
	\widetilde{\mathbf{Y}}_{i,u}^{\rm BS}=\mathbf{F}_i{\rm diag}(\mathbf{h}_{i,u})\mathring{\mathbf{V}}_i+\widetilde{\mathbf{N}}^{\rm BS}_{i,u} 
	\approx\mathbf{A}_{\rm  B}\bm{\Xi}_i\mathbf{A}_{\rm L_i}^H{\rm diag}(\mathbf{h}_{i,u})\mathring{\mathbf{V}}_i+\widetilde{\mathbf{N}}^{\rm BS}_{i,u},
	\setlength{\belowdisplayskip}{3.2pt}
\end{equation}
where $\mathring{\mathbf{V}}_i$ is the same as the previous definition, $\widetilde{\mathbf{N}}^{\rm BS}_{i,u}=[\widetilde{\mathbf{n}}_{i,u,1}^{\rm BS},\cdots,\widetilde{\mathbf{n}}_{i,u,Q_i}^{\rm BS}]\in\mathbb{C}^{J\times Q_i}$.

Assuming that $\widehat{\mathbf{h}}_{i,u}$ is the estimated channel, we denote $\mathbf{A}_{\rm L_i}^H{\rm diag}(\widehat{\mathbf{h}}_{i,u})\mathring{\mathbf{V}}_i$ by $\bm{\Phi}_{i,u}\in\mathbb{C}^{G_r\times Q_i}$.
As stated before, the common channel $\mathbf{F}$ can be estimated by joint processing of multi-user signals.
By collecting all users' signals, we attain
\begin{equation}\label{Yddot}
	\setlength{\abovedisplayskip}{3.2pt}
	\mathbf{\ddot{Y}}^{\rm BS}_i\approx\mathbf{A}_{\rm B}\bm{\Xi}_i\bm{\ddot{\Phi}}_i+\mathbf{\ddot{N}}_i,
	\setlength{\belowdisplayskip}{3.2pt}
\end{equation}
where $\mathbf{\ddot{Y}}^{\rm BS}_i=[\widetilde{\mathbf{Y}}_{i,1}^{\rm BS},\cdots,\widetilde{\mathbf{Y}}_{i,U}^{\rm BS}]\in\mathbb{C}^{J\times Q_iU}$, $\bm{\ddot{\Phi}}_i=[\bm{\Phi}_{i,1},\cdots,\bm{\Phi}_{i,U}]\in\mathbb{C}^{G_r\times Q_iU}$ and $\mathbf{\ddot{N}}_i=[\widetilde{\mathbf{N}}^{\rm BS}_{i,1},\cdots,\widetilde{\mathbf{N}}^{\rm BS}_{i,U}]\in\mathbb{C}^{J\times Q_iU}$.
According to the Kronecker CS framework, vectorizing $\mathbf{\ddot{Y}}^{\rm BS}_i$ yields
\begin{equation}\label{CS_joint}
	\setlength{\abovedisplayskip}{3.2pt}
	\mathbf{\ddot{y}}_i^{\rm BS}={\rm vec}(\mathbf{\ddot{Y}}^{\rm BS}_i)\approx (\bm{\ddot{\Phi}}_i^T\otimes\mathbf{A}_{\rm B})\bm{\xi}_i + {\rm vec} (\mathbf{\ddot{N}}_i),
	\setlength{\belowdisplayskip}{3.2pt}
\end{equation}
where $\bm{\xi}_i={\rm vec}(\bm{\Xi}_i)\in\mathbb{C}^{G_tG_r\times 1}$ and $\bm{\ddot{\Phi}}_i^T\otimes\mathbf{A}_{B}\in\mathbb{C}^{Q_iUJ\times G_tG_r}$. 
 Since the vectors in the sensing matrix $\bm{\ddot{\Phi}}_i^T\otimes\mathbf{A}_{\rm B}$ created by the Kronecker product are excessively lengthy, direct processing of Eqn. (\ref{CS_joint}) will result in high computing cost regardless of the recovery algorithm used.
In the later section, we will offer a fast solution for Eqn. (\ref{Yddot}) that avoids the Kronecker product.

\subsubsection{Estimation of $\mathbf{D}$}
Here, both RISs are turned on simultaneously to estimate $\mathbf{D}$ based on the estimated $\{\widehat{\mathbf{F}}_1,\widehat{\mathbf{F}}_2\}$. By removing the received signals related to the two RISs' cascaded channels, the remaining signal at the BS in the $q_3$-th sub-frame is approximated as
$
\mathbf{y}^{\rm BS}_{q_3,u}\approx\widehat{\mathbf{F}}_2\mathbf{V}_{2,q_3}\mathbf{D}\mathbf{V}_{1,q_3}\widehat{\mathbf{h}}_{1,u}+\mathbf{N}^{\rm BS}_{q_3,u}\mathbf{s}_u, \ \forall u$.
According to the proposed MAE strategy, we jointly process all users' signals to estimate $\mathbf{D}$. By collecting all users' signals in the $q_3$-th sub-frame, we have $
	\overline{\mathbf{Y}}^{\rm BS}_{q_3}=\widehat{\mathbf{F}}_2\mathbf{V}_{2,q_3}\mathbf{D}\mathbf{V}_{1,q_3}\widehat{\mathbf{H}}_{1}+\overline{\mathbf{N}}^{\rm BS}_{q_3}$,
where $\widehat{\mathbf{H}}_1=[\widehat{\mathbf{h}}_{1,1},\cdots,\widehat{\mathbf{h}}_{1,U}]\in\mathbb{C}^{L\times U}$ and $\overline{\mathbf{N}}^{\rm BS}_{q_3}=[\mathbf{N}^{\rm BS}_{q_3,1}\mathbf{s}_1,\cdots,\mathbf{N}^{\rm BS}_{q_3,U}\mathbf{s}_U]\in\mathbb{C}^{J\times U}$. 

Following that, we divide the total sub-frames $Q_3$ into $N_X$ and $N_Y$ ($Q_3=N_X\times N_Y$) for phase adjustment of RIS 1 and 2, respectively.
Then the collected signal in $Q_3$ sub-frames $\overline{\mathbf{Y}}^{\rm BS}\in\mathbb{C}^{N_YJ\times N_XU}$ is written as
\begin{equation}\label{DBS1}
	\setlength{\abovedisplayskip}{3.2pt}
	\begin{aligned}
	\overline{\mathbf{Y}}^{\rm BS}&=
\left[\mathbf{V}_{2,1}^T\widehat{\mathbf{F}}_2^T,\cdots,\mathbf{V}_{2,N_Y}^T\widehat{\mathbf{F}}_2^T\right]^T\mathbf{D}
\begin{bmatrix}
\mathbf{V}_{1,1}\widehat{\mathbf{H}}_{1},\cdots,\mathbf{V}_{1,N_X}\widehat{\mathbf{H}}_{1}
\end{bmatrix} 
+\overline{\mathbf{N}}^{\rm BS}\\
&\approx
\left[\mathbf{V}_{2,1}^T\widehat{\mathbf{F}}_2^T,\cdots,\mathbf{V}_{2,N_Y}^T\widehat{\mathbf{F}}_2^T\right]^T	\mathbf{A}_{\rm L_2}\bm{\Delta}\mathbf{A}_{\rm L_1}^H
\begin{bmatrix}
	\mathbf{V}_{1,1}\widehat{\mathbf{H}}_{1},\cdots,\mathbf{V}_{1,N_X}\widehat{\mathbf{H}}_{1}
\end{bmatrix}+\overline{\mathbf{N}}^{\rm BS},
\end{aligned}
	\setlength{\belowdisplayskip}{3.2pt}
\end{equation}
where $\overline{\mathbf{N}}^{\rm BS}\in\mathbb{C}^{N_YJ\times N_XU}$ is the noise matrix.
Via kronecker CS, the above equation can be converted to
\begin{equation}\label{DBS2}
		\setlength{\abovedisplayskip}{3.2pt}
\bar{\mathbf{y}}^{\rm BS}={\rm vec}(\overline{\mathbf{Y}}^{\rm BS})\approx(\mathbf{A}_{\rm H}^T\otimes\mathbf{A}_{\rm F})\bm{\delta}+{\rm vec}(\overline{\mathbf{N}}^{\rm BS}),
\setlength{\belowdisplayskip}{3.2pt}
\end{equation}
where $\mathbf{A}_{\rm H}=\mathbf{A}_{\rm L_1}^H
\begin{bmatrix}
	\mathbf{V}_{1,1}\widehat{\mathbf{H}}_{1},\cdots,\mathbf{V}_{1,N_X}\widehat{\mathbf{H}}_{1}
\end{bmatrix}\in\mathbb{C}^{G_r\times N_XU}$,  $\mathbf{A}_{\rm F}=\begin{bmatrix}
\mathbf{A}_{\rm L_2}^T\mathbf{V}_{2,1}^T\widehat{\mathbf{F}}_2^T,\cdots,\mathbf{A}_{\rm L_2}^T\mathbf{V}_{2,N_Y}^T\widehat{\mathbf{F}}_2^T
\end{bmatrix}^T\in\mathbb{C}^{N_YJ\times G_r}$ and $\bm{\delta}={\rm vec}(\bm{\Delta})\in\mathbb{C}^{G_r^2\times 1}$.

Evidently, the kronecker product imposes the same computing burden on Eqn. (\ref{DBS2}) as Eqn. (\ref{CS_joint}). As indicated previously, this requires a computationally efficient framework for estimating several parameters $\theta^{\rm ele},\theta^{\rm azi},\phi^{\rm ele},\phi^{\rm azi}$ and $\beta$ inherent in Eqn. (\ref{D}), which will be presented in the later section.
\subsection{Small-Timescale Estimation}\label{smallscale}

As the channel $\mathbf{F}_i$ known, channel estimates of $\{\mathbf{h}_{i,u}\}_{u=1}^U$ can be performed at the BS with little training overhead, as contrast to the case of Eqn. (\ref{RIS_yu}).

Recalling Eqn. (\ref{YBSi}), the BS's collected signal reflected by RIS $i$ in $\bar{Q}_i$ sub-frames can be given by 
\begin{equation}\label{T2}
	\setlength{\abovedisplayskip}{3.2pt}
	\begin{aligned}
 	\begin{bmatrix}
 		\widetilde{\mathbf{y}}^{\rm BS}_{i,u,1}\\
 		\vdots\\
 		\widetilde{\mathbf{y}}^{\rm BS}_{i,u,\bar{Q}_i}
 	\end{bmatrix}=&
\begin{bmatrix}
	\widehat{\mathbf{F}}_i\mathbf{V}_{i,1} \\
	\vdots\\
	\widehat{\mathbf{F}}_i\mathbf{V}_{i,\bar{Q}_i}
\end{bmatrix}
 \mathbf{h}_{i,u}+\widetilde{\mathbf{n}}_{i,u}^{\rm BS}
 \approx\bm{\Psi}\bm{\zeta}_{i,u}+\widetilde{\mathbf{n}}_{i,u}^{\rm BS},
\end{aligned}
\setlength{\belowdisplayskip}{3.2pt}
\end{equation}
where $\bm{\Psi}=\begin{bmatrix}
	\mathbf{A}_{\rm L_i}^T\mathbf{V}_{i,1}^T\widehat{\mathbf{F}}_i^T,\cdots,\mathbf{A}_{\rm L_i}^T\mathbf{V}_{i,\bar{Q}_i}^T\widehat{\mathbf{F}}_i^T
\end{bmatrix}^T\in\mathbb{C}^{\bar{Q}_iJ\times G_r}$ is the sensing matrix. Compared to the sensing matrix used to recover $\bm{\zeta}_{i,u}$ in Eqn. (\ref{RIS_yu}), the number of measurements of $\bm{\Psi}$ is greater owing to the multiplication by the number of BS antennas. This implies that training costs for estimating $\mathbf{h}_{i,u}$ can be significantly reduced if $\mathbf{F}_i$ is known, i.e., $\bar{Q}_i\ll Q_i$.
\section{Low-Complexity Bayesian Compressive Sensing Solution}\label{S4}

This section offers low-complexity Bayesian learning-based solutions to the problems outlined in the preceding section. One is the basic MAE-based CS issue for the RIS-user channel, like Eqn. (\ref{RIS_yu}) and (\ref{T2}), while the other is a more complicated CS issue in the face of more recovery parameters, as Eqn. (\ref{Yddot}) and (\ref{DBS1}). 
We will focus on the latter problem,
since the former problem can be addressed using the off-grid single measurement vector (SMV)-based EM-GAMP method, which is a simplified version of the latter problem's solution.

\subsection{LoS-Aided Dictionary Development}\label{LOS}
Let's give the necessity of the LoS-aided dictionary first.
 Considering a one-dimensional DFT bin, i.e.,  $\{-1+\frac{2g}{G}|g=0,\cdots,G-1\}$, when we sample multiple true angles $\sin(x)$ in this bin, they will be impacted by the mismatch problem due to the limited bin size $G$. Although off-grid techniques can overcome this problem, the algorithmic error is inevitable.
 If we know one of the true angles $\overline{x}$, we can design a $\overline{x}$-based sampling method, i.e., 
\begin{equation}\label{dic}
	\setlength{\abovedisplayskip}{3.2pt}
	\sin(x)=\begin{cases}
		\overline{x}+\frac{2g}{G}, & 	\overline{x}+\frac{2g}{G}\leq 1, \\ -2+\overline{x}+\frac{2g}{G}, & 	\overline{x}+\frac{2g}{G}> 1,
	\end{cases} \ g=1,\cdots,G-1.
\setlength{\belowdisplayskip}{3.2pt}
\end{equation}
For a UPA dictionary $\mathbf{A}_x$ with the known LoS path's angle $\overline{x}^{\rm ele/azi}$, it can be constructed by
\begin{equation}\label{Ax}
	\setlength{\abovedisplayskip}{3.2pt}
\mathbf{A}_x=
\{\mathbf{a}(\sin(x^{\rm ele}_{g_z}),\sin(x^{\rm azi}_{g_y}))|g_z=1,\cdots,G_z,g_y=1,\cdots,G_y\},
\setlength{\belowdisplayskip}{3.2pt}
\end{equation}
 where $\sin(x^{\rm ele}_{g_z})$ and $\sin(x^{\rm azi}_{g_y})$ are determined by Eqn. (\ref{dic}) with $\{G_z,\overline{x}^{\rm ele}\}$ and $\{G_y,\overline{x}^{\rm azi}\}$, respectively.

According to the locations of the BS, RIS 1 and RIS 2, the LoS path's angle $\overline{\vartheta}^{\rm ele/azi}_{i}$, $\overline{\varphi}^{\rm ele/azi}_{i}$, $\overline{\theta}^{\rm ele/azi}$, and $\overline{\phi}^{\rm ele/azi}$ are easily obtained. When recovering $\{{\vartheta}^{\rm ele/azi}_{p,1/2},{\varphi}^{\rm ele/azi}_{p,1/2},{\theta}^{\rm ele/azi}_b,{\phi}^{\rm ele/azi}_b|p=1,\cdots, P_{f_i},b=1\cdots,P_d\}$, the corresponding dictionary can be established as above. Thus, the recovery of the LoS path will not be affected by the dictionary size. Along with this, the off-grid procedure is simplified since the on-grid LoS path's parameters have been optimal.
In particular, the dictionary of the BS/RIS may be different when representing different channels. For instance, $\mathbf{A}_{\rm L_2}$  depends on $\overline{\varphi}$ for $\mathbf{F}_2$ but $\overline{\theta}$ for $\mathbf{D}$.

\subsection{SVD-MMV-CS Framework}\label{SVDMCS}
 Eqn. (\ref{Yddot}) and (\ref{DBS1}) belong to the same class of problems, we shall derive the formula based on Eqn. (\ref{DBS1}) without losing generality.
As described previously, we have
$
\overline{\mathbf{Y}}^{\rm BS}\approx
\mathbf{A}_{\rm F}\mathbf{A}_{\rm L_2}\bm{\Delta}\mathbf{A}_{\rm L_1}^H\mathbf{A}_{\rm H}+\overline{\mathbf{N}}^{\rm BS}.
$
Via a rank decomposition of $\bm{\Delta}=\sum_{k=1}^{r_{\Delta}}\mathbf{d}_{1,k}\mathbf{d}_{2,k}^H$, with ${\rm rank}(\bm{\Delta})=r_{\Delta}$, we can obtain
\begin{equation}\label{DY}
	\setlength{\abovedisplayskip}{3.2pt}
\overline{\mathbf{Y}}^{\rm BS}\approx\sum_{k=1}^{r_{\Delta}}(\mathbf{A}_{\rm F}\mathbf{A}_{\rm L_2}\mathbf{d}_{1,k})(\mathbf{A}_{\rm H}^H\mathbf{A}_{\rm L_1}\mathbf{d}_{2,k})^H+\overline{\mathbf{N}}^{\rm BS}.
\setlength{\belowdisplayskip}{3.2pt}
\end{equation}

Note that $\bm{\Delta}$ is a matrix with $P_d$-sparsity; hence, ${\rm rank}(\overline{\mathbf{Y}}^{\rm BS})\leq{\rm rank}(\bm{\Delta})=r_{\Delta}\leq P_d$. We denote $\mathcal{C}(\bm{\Delta})$ by the column space of $\bm{\Delta}$, which is the vector subspace whose elements are $P_d$-sparse, as well as $\mathcal{C}(\bm{\Delta}^H)$.
Notably, we assume that the AoA and AoD are one-to-one mapping and the sensing matrices meet the null space property (NSP) \cite{NSP}, then
${\rm dim}(\mathbf{A}_{\rm F}\mathbf{A}_{\rm L_2}\mathcal{C}(\bm{\Delta}))={\rm dim}(\mathbf{A}_{\rm L_i}^H\mathbf{A}_{\rm H}\mathcal{C}(\bm{\Delta}^H))={\rm rank}(\bm{\Delta})=r_{\Delta}$. Thus, the decomposition of 
$\overline{\mathbf{Y}}^{\rm BS}$ in Eqn. (\ref{DY}) is a $\rm rank$-$r_{\Delta}$ decomposition due to $\mathbf{d}_{1,k}\in\mathcal{C}(\bm{\Delta})$ and $\mathbf{d}_{2,k}\in\mathcal{C}(\bm{\Delta}^H)$.
 According to the above, the following proposition can be developed.

\textbf{Theorem 1:} Let two sensing matrices be $\bm{\Phi}_1\in\mathbb{C}^{N_1\times M_1}$ and $\bm{\Phi}_2\in\mathbb{C}^{N_2\times M_2}$, respectively. If $\bm{\Delta}\in\mathbb{C}^{M_1\times M_2}$ is $P$-sparse, and $\mathbf{Y}$ w.r.t. $\bm{\Delta}=\sum_{k=1}^{{\rm rank}(\bm{\Delta})}\mathbf{d}_{1,k}\mathbf{d}_{2,k}^H$ can be given by the form of Eqn. (\ref{DY}), i.e.,
\begin{equation}\label{YPHI}
	\setlength{\abovedisplayskip}{3.2pt}
\mathbf{Y}=\bm{\Phi}_1\bm{\Delta}\bm{\Phi}_2^H=\sum_{k=1}^{{\rm rank}(\bm{\Delta})}(\bm{\Phi}_1\mathbf{d}_{1,k})(\bm{\Phi}_2\mathbf{d}_{2,k})^H\in\mathbb{C}^{N_1\times N_2},
\setlength{\belowdisplayskip}{3.2pt}
\end{equation}
 then $\bm{\Delta}$ can be recovered uniquely via rank decomposition, such as SVD. We assume
$
	\mathbf{Y}=(\mathbf{\ddot{U}}\sqrt{\bm{\Sigma}})(\sqrt{\bm{\Sigma}}\mathbf{\ddot{V}}^H)=\sum_{k=1}^{{\rm rank}(\mathbf{Y})}\mathbf{e}_{1,k}\mathbf{e}_{2,k}^H
$,
where $\mathbf{\ddot{U}}$, $\mathbf{\ddot{V}}$ are unitary matrices, and $\bm{\Sigma}$ is the singular value matrix.
Let $\hat{\bm{\delta}}_{i,k}\in\mathbb{C}^{M_j\times 1}$, in which $i\in\{1,2\}$ and $k\in\{1,2,\cdots,{\rm rank} (\mathbf{Y})\}$, be the solution of $ {\rm min}\{ \Vert\bm{\delta}_{i,k} \Vert_0+ \Vert\mathbf{e}_{i,k}-\bm{\Phi}_i\bm{\delta}_{i,k}\Vert_2^2\}$. Since each $\bm{\delta}_{i,k}$ is $P$-sparse, MMV issues can be naturally formulated as follows. Re-writing that $\mathbf{Y}=\mathbf{E}_1\mathbf{E}_2^H$ with $\mathbf{E}_1=[\mathbf{e}_{1,k},\cdots,\mathbf{e}_{1,{\rm rank}(\mathbf{Y})}]\in\mathbb{C}^{N_1\times{\rm rank}(\mathbf{Y})}$ and $\mathbf{E}_2=[\mathbf{e}_{2,k},\cdots,\mathbf{e}_{2,{\rm rank}(\mathbf{Y})}]\in\mathbb{C}^{N_2\times {\rm rank}(\mathbf{Y})}$. For $i\in\{1,2\}$, the MMV issue is
\begin{equation}\label{E}
	\setlength{\abovedisplayskip}{3.2pt}
	\underset{\mathbf{\bm{\Delta}}_{i}^{\rm M}}{\rm arg \ min} \ \Vert\bm{\Delta}_{i}^{\rm M} \Vert_0, \
	 {\rm s.t.} \  \Vert\mathbf{E}_{i}-\bm{\Phi}_i\bm{\Delta}_{i}^{\rm M}\Vert_F^2\leq \epsilon, 
{\rm supp}(\mathbf{t}_{i,1})=	\cdots={\rm supp}(\mathbf{t}_{i,R}),
\setlength{\belowdisplayskip}{3.2pt}
\end{equation}
where $\bm{\Delta}_{i}^{\rm M}=[\mathbf{t}_{i,1},\cdots,\mathbf{t}_{i,R}]\in\mathbb{C}^{{N_i}\times R}$ is the matrix with group sparsity, $R={\rm rank}({\mathbf{Y}})$,
and ${\rm supp}(\cdot)$ denotes the sparsity support.
Finally, the recovered signal can be written as
\begin{equation}\label{DD}
	\setlength{\abovedisplayskip}{3.2pt}
	\widehat{\bm{\Delta}}=	\widehat{\bm{\Delta}}^{\rm M}_1	(\widehat{\bm{\Delta}}^{{\rm M}}_2)^H.
\setlength{\belowdisplayskip}{3.2pt}
\end{equation}

\textbf{Proof:} See Appendix \ref{appendixA}.

Based on $\textbf{Theorem 1}$, the recovery issue of Eqn. (\ref{DBS1}) is solved as follows. First, decomposing $\overline{\mathbf{Y}}^{\rm BS}$ into $\sum_{k=1}^{{\rm rank}(\overline{\mathbf{Y}}^{\rm BS})}\mathbf{e}_{1,k}\mathbf{e}_{2,k}^H$ by SVD. Next, stacking $\{\mathbf{e}_{i,k}\}_{k=1}^{{\rm rank}(\overline{\mathbf{Y}}^{\rm BS})}$ into $\mathbf{E}_i$. Subsequently, using $\bm{\Phi}_1=\mathbf{A}_{\rm F}$ and $\bm{\Phi}_2=\mathbf{A}_{\rm H}$ to recover $\bm{\Delta}_i^{\rm M}$ according to Eqn. (\ref{E}). Finally, the sparse signal $\bm{\Delta}$ in Eqn. (\ref{DBS1}) is estimated by $	\widehat{\bm{\Delta}}=	\widehat{\bm{\Delta}}^{\rm M}_1	(\widehat{\bm{\Delta}}^{{\rm M}}_2)^H$.

\subsection{M-GAMP}\label{M-GAMP}


Similar to the GAMP algorithm, M-GAMP consists of the two main components: prior distribution setup and posterior distribution approximation. According to $\textbf{Theorem 1}$, the whole recovery problem in Eqn. (\ref{DBS1}) is decoupled into two separate subproblems as Eqn. (\ref{E}). Due to the independence of these two identical subproblems, we ignore the subscript $i$ to obtain 
\begin{equation}\label{E2}
	\setlength{\abovedisplayskip}{3.2pt}
		\underset{\mathbf{\bm{\Delta}}^{\rm M}}{\rm arg \ min} \ \Vert\bm{\Delta}^{\rm M} \Vert_0+ \Vert\mathbf{E}-\bm{\Phi}\bm{\Delta}^{\rm M}\Vert_F^2, 
		 \ \ \  {\rm s.t.} \ {\rm supp}(\mathbf{t}_{1})=	\cdots={\rm supp}(\mathbf{t}_{R}).
		 \setlength{\belowdisplayskip}{3.2pt}
\end{equation}
We assume that $\mathbf{E}\in\mathbb{C}^{M\times R}$, $\bm{\Phi}\in\mathbb{C}^{M\times G}$ and $\mathbf{\bm{\Delta}}^{\rm M}=[\mathbf{t}_1,\cdots,\mathbf{t}_{R}]\in\mathbb{C}^{G\times R}$.
\subsubsection{Prior Distribution Setup}
To characterize the channel's sparsity, the Gaussian mixture model (GMM)-based prior distribution is adopted, which is the extension of the Bernoulli-Gaussian distribution.
Assume that the elements in $\bm{\Delta}^{\rm M}$ follow the i.i.d. Gaussian mixture distribution, i.e.,
\begin{equation}\label{prior}
	\setlength{\abovedisplayskip}{3.2pt}
	p(t_{g,r})=(1-\kappa_{g})\delta(t_{g,r})+\kappa_{g}\sum_{l=1}^{L}\omega_{r,l}\mathcal{CN}(t_{g,r}|\nu_{r,l},\varsigma_{r,l}),
	\setlength{\belowdisplayskip}{3.2pt}
\end{equation}
where $\delta(\cdot)$ is the Dirac function, $\kappa_{g}$ is the common sparsity rate for $\forall r$. $\omega_{r,l}$, $\nu_{r,l}$ and $\varsigma_{r,l}$ denote the weight coefficient, mean and variance of the $l$-th GM component, respectively. In particular, the weight coefficients satisfy $\sum_{l=1}^{L}\omega_{l}=1$.
Additionally, we suppose that the noise follows the complex Gaussian distribution with variance $\rho$, i.e., $
	p(\mathring{n}_{m,r}|\rho_r)=\mathcal{CN}(\mathring{n}_{m,r}|0,\rho_r)$.
Finally, let $\bm{\Theta}=\{\kappa_g,\omega_{r,l},\mu_{r,l},\varsigma_{r,l},\rho_r|\forall g,r,l\}$ be the hyperparameter set mentioned above.
\subsubsection{Posterior Distribution Approximation via M-GAMP }
 In the construction of the GAMP algorithm, $p(\eta_{m,r}|\mathbf{E})$  and $p(t_{g,r}|\mathbf{E})$  approximations are adopted. 
\begin{algorithm}[!t] 
	\caption{Off-Grid M-EM-GAMP } 
	\label{M-EM-GAMP}      
	\begin{algorithmic}[1] 
		\footnotesize{
			\REQUIRE { Signal $\mathbf{Y}\in\mathbb{C}^{M\times R}$, sensing matrix $\bm{\Phi}\in\mathbb{C}^{M\times G}$, and iteration number $Z$, $O_{\rm EM}$ and $\mathcal{T}$.
			}

			\ENSURE {Estimated sparse signal $\widehat{\bm{\Delta}}^{\rm M}$.} 
			
			\STATE{$\textbf{Initialize:}$  $z=o=t=1$,  $\forall m,r:\hat{s}_{m,r}(0)=0$. Given a random $\bm{\Theta}^o$, $\{\{\hat{t}_{g,r}(z)\}_{q=1}^{G}\}_{r=1}^R$ and $\{\{\tau_{g,r}^{t}(z)\}_{g=1}^{G}\}_{r=1}^R$ are initialized as the mean and variance of the prior distribution in Eqn. (\ref{prior}).
			}

			\STATE{\textbf{\%\% M-EM-GAMP Procedure}}
			\REPEAT	
			\STATE{\emph{Expectation-Step:}}	
		
			\REPEAT
			\STATE{\emph{Output linear step}:}
			\STATE{\ \  $\forall m,r:\ \tau_{m,r}^{\mu}(z)=\sum_{g=1}^{G}|[\bm{\Phi}]_{m,g}|^2\tau_{g,r}^{t}(z)$}
			\STATE{\ \   $\forall m,r:\ \hat{\mu}_{m,r}(z)=\sum_{g=1}^{G}[\bm{\Phi}]_{m,g}\hat{t}_g(z)-\tau_m^\mu(z)\hat{s}_m(z-1)$}
			
			\STATE{\emph{Output nonlinear step}:}
			\STATE{\ \ $\forall {m,r}:\
				\hat{s}_{m,r}(z)=g_{out}(\hat{\mu}_{m,r}(z),\tau_{m,r}^{\mu}(z),\bm{\Theta}^o)$}
			\STATE{\ \   $\forall m,r:\
				\tau_{m,r}^s(z)=-\frac{\partial}{\partial\hat{\mu}_m}g_{out}(\hat{\mu}_{m,r}(z),\tau_{m,r}^{\mu}(z),\bm{\Theta}^o)$}
			\STATE{\emph{Input linear step}:}
			\STATE{\ \   $\forall g,r:\
				\tau_{g,r}^\psi(z)=\left( \sum_{m=1}^{M}|[\bm{\Phi}]_{m,g}|^2\tau_{m,r}^s(z)\right)^{-1}$}
			\STATE{\ \  $\forall g,r:\
				\hat{\psi}_{g,r}(z)=\hat{t}_{g,r}(z)+\tau_{g,r}^\psi(z)\sum_{m=1}^{M}[\bm{\Phi}]_{m,g}\hat{s}_{m,r}(z)$}
			\STATE{\emph{Input nonlinear step}:}
			\STATE{\ \   $\forall g,r:\
				\hat{t}_{g,r}(z+1)=g_{in}(\hat{\psi}_{g,r}(z),\tau_{g,r}^\psi(z),\bm{\Theta}^o)$}
			\STATE{\ \   $\forall g,r:\
				\tau^{t}_{g,r}(z+1)=\tau_{g,r}^\psi(z)\frac{\partial}{\partial \hat{\psi}_{g,r}  }g_{in}(\hat{\psi}_{g,r}(z),\tau_{g,r}^\psi(z),\bm{\Theta}^o) $}
			\UNTIL{ $z+1\geq Z$ or $\sum_{g=1}^{G}\sum_{r=1}^{R}\vert\hat{t}_{g,r}(z+1)-\hat{t}_{g,r}(z)\vert^2<\epsilon$ is reached.}
		
			\STATE{\emph{Maximization-Step:}}
	
			\STATE{ $\forall r$: Update $\rho_r^{o+1}$ according to Eqn. (\ref{rho}). }
			\FOR{$l=1,\cdots,L$}
			\STATE{$\forall r$: Update $\nu_{r,l}^{o+1}$, $\varsigma_{r,l}^{o+1}$ and $\omega_{r,l}^{o+1}$ according to Eqn. (\ref{nu}),  (\ref{varsig}) and (\ref{xi}).}
			
			\ENDFOR
			\STATE{$\forall g$: Update $\kappa^{o+1}_{g}$ according to Eqn. (\ref{kappa}).}
		
			\UNTIL{$o+1\geq O_{\rm EM}$ is reached.}
			\STATE{\textbf{\%\% Off-Grid Refinement}}
		}
		\STATE{ $\{\widetilde{\theta}^{{\rm ele},1}_b\}_{b=1}^{P_{h,u}}$ and $\{\widetilde{\theta}^{{\rm azi},1}_b\}_{b=1}^{P_{h,u}}$ are initialized by the above solution of Eqn. (\ref{E2}).}
		\REPEAT
		\STATE{
			Compute $\{\widetilde{\theta}^{{\rm ele},t+1}_b\}_{b=1}^{P_{h,u}}$ and $\{\widetilde{\theta}^{{\rm azi},t+1}_b\}_{b=1}^{P_{h,u}}$ according to Eqn. (\ref{thetat1})-(\ref{theta2}).
		}
	\STATE{ Compute path gains according to Eqn. (\ref{DLS}) and update the residual according to Eqn. (\ref{DRES}).
	}
		\UNTIL{$t+1\geq\mathcal{T}$ or the stop condition is reached.}
	\STATE{$\widehat{\bm{\Delta}}^{\rm M}$ is obtained by filling zero rows into $\widetilde{\bm{\Delta}}^{{\rm M},\mathcal{T}}$.}
	\end{algorithmic}
\end{algorithm}

Firstly, to establish the relationship between the noiseless output and the measurement matrix $\mathbf{E}$, we have ${\eta}_{m,r}=[\bm{\Phi}]_{m,:}[\bm{\Delta}^{\rm M}]_{:,r}$. The true posterior $p(\eta_{m,r}|\mathbf{E},\bm{\Theta})$ is approximated by GAMP:
\begin{equation}\label{PZY}
	\setlength{\abovedisplayskip}{3.2pt}
\hat{p}(\eta_{m,r}|\mathbf{E},\bm{\Theta},\hat{\mu}_{m,r},\tau_{m,r}^{\mu})=\frac{p(e_{m,r}|\eta_{m,r},\bm{\Theta})\mathcal{CN}(\eta_{m,r}|\hat{\mu}_{m,r},\tau_{m,r}^{\mu})}{\int_{\eta_{m,r}}p(e_{m,r}|\eta_{m,r},\bm{\Theta})\mathcal{CN}(\eta_{m,r}|\hat{\mu}_{m,r},\tau_{m,r}^{\mu})},
\setlength{\belowdisplayskip}{3.2pt}
\end{equation}
where $\hat{\mu}_{m,r}$ and $\tau_{m,r}^{\mu}$ denote the mean and variance of $\eta_{m,r}$, respectively, which are iteratively updated during the GAMP procedure. Under the additive white Gaussian noise assumption, we obtain $p(e_{m,r}|\eta_{m,r},\bm{\Theta})=\mathcal{CN}(e_{m,r}|\eta_{m,r},\rho)$. Further, the mean and variance of Eqn. (\ref{PZY}) can be expressed as
\begin{equation}
	\setlength{\abovedisplayskip}{3.2pt}
	\hat{\eta}_{m,r}=\hat{\mu}_{m,r}+\frac{\tau_{m,r}^\mu}{\tau_{m,r}^\mu+\rho}(e_{m,r}-\hat{\mu}_{m,r}),
\
	{\tau}_{m,r}^{\eta}=\frac{\tau_{m,r}^\mu\rho}{\tau_{m,r}^\mu+\rho},
	\setlength{\belowdisplayskip}{3.2pt}
\end{equation}
where $\hat{\eta}_{m,r}$ and ${\tau}_{m,r}^{\eta}$ are iteratively updated during the GAMP procedure.

Since GAMP assumes posterior independence among hidden variables $\{\{{t}_{g,r}\}_{g=1}^{G}\}_{r=1}^R$, it approximates the true posterior $p(t_{g,r}|\mathbf{E},\bm{\Theta})$ by
\begin{equation}\label{PXY}
	\setlength{\abovedisplayskip}{3.2pt}
	\hat{p}(t_{g,r}|\mathbf{E},\bm{\Theta},\hat{\psi}_{g,r},\tau_{g,r}^{\psi})=\frac{p(t_{g,r}|\bm{\Theta})\mathcal{CN}(t_{g,r}|\hat{\psi}_{g,r},\tau_{g,r}^{\psi})}{\int_{t_{g,r}}p(t_{g,r}|\bm{\Theta})\mathcal{CN}(t_{g,r}|\hat{\psi}_{g,r},\tau_{g,r}^{\psi})},
	\setlength{\belowdisplayskip}{3.2pt}
\end{equation}
 Using the multiplication rule of the Gaussian distribution, i.e.,
\begin{equation}
	\setlength{\abovedisplayskip}{3.2pt}
\mathcal{CN}(x|\psi_a,\tau_a) \mathcal{CN}(x|\psi_b,\tau_b)= \mathcal{CN}(x|\frac{\mu_a/\tau_a+\psi_b/\tau_b}{1/\tau_a+1/\tau_b},\frac{1}{1/\tau_a+1/\tau_b})
		\mathcal{CN}(x|\psi_a-\psi_b,\tau_a+\tau_b),
		\setlength{\belowdisplayskip}{3.2pt}
\end{equation}
Eqn. (\ref{PXY}) with Eqn. (\ref{prior}) substituted is given by
\begin{equation}
	\setlength{\abovedisplayskip}{3.2pt}
	\begin{aligned}
		&\hat{p}(t_{g,r}|\mathbf{E},\bm{\Theta},\hat{\psi}_{g,r},\tau_{g,r}^{\psi})\\
		&=\frac{(1-\kappa_g)\delta(t_{g,r})\mathcal{CN}(t_{g,r}|\hat{\psi}_{g,r},\tau_{g,r}^{\psi})}{		\int_{t_{g,r}}p(t_{g,r}|\bm{\Theta})\mathcal{CN}(t_{g,r}|\hat{\psi}_{g,r},\tau_{g,r}^{\psi})}   +\frac{		\kappa_g\sum_{l=1}^{L}\omega_{l}\mathcal{CN}(t_{g,r}|\nu_{l},\varsigma_{l})\mathcal{CN}(t_{g,r}|\hat{\psi}_{g,r},\tau_{g,r}^{\psi})}{\int_{t_{g,r}}p(t_{g,r}|\bm{\Theta})\mathcal{CN}(t_{g,r}|\hat{\psi}_{g,r},\tau_{g,r}^{\psi})}
		\\
		&=(1-\varpi_{g,r})\delta(t_{g,r})+\varpi_{g,r}\sum_{l=1}^{L}\upsilon_{g,r,l}\mathcal{CN}(t_{g,r}|\varrho_{g,r,l},\varkappa_{g,r,l}) 
	\end{aligned}
\setlength{\belowdisplayskip}{3.2pt}
\end{equation}
with $	\varrho_{g,r,l}=\frac{\hat{\psi}_{g,r}/\tau_{g,r}^\psi+\nu_{r,l}/\varsigma_{r,l}}{1/\tau_{g,r}^\psi+1/\varsigma_{r,l}}$, $	\varkappa_{g,r,l}=\frac{1}{1/\tau_{g,r}^\psi+1/\varsigma_{r,l}}$, and
\begin{equation}
	\setlength{\abovedisplayskip}{3.2pt}
	\varpi_{g,r}=\frac{1}{1+\left(\frac{\sum_{l=1}^{L}\rho \omega_{r,l}\mathcal{CN}(\hat{\psi}_{g,r}|\nu_{r,l}, \varsigma_{r,l}+\tau_{g,r}^\psi)}{(1-\rho)\mathcal{CN}(0|\hat{\psi}_{g,r},\tau_{g,r}^\psi)}\right)^{-1}},
	\setlength{\belowdisplayskip}{3.2pt}
\end{equation}
 \begin{equation}
 	\setlength{\abovedisplayskip}{3.2pt}
\upsilon_{g,r,l}=\frac{\rho \omega_{r,l}\mathcal{CN}(\hat{\psi}_{g,r}|\nu_{r,l}\varsigma_{r,l}+\tau_{g,r}^\psi)}{\sum_{l=1}^{L}\rho \omega_i\mathcal{CN}(\hat{\psi}_{g,r}|\nu_{r,l},\varsigma_{r,l}+\tau_{g,r}^\psi)}.
\setlength{\belowdisplayskip}{3.2pt}
\end{equation}

\subsubsection{Input and Output Functions}
With the above approximations, two scalar functions are defined: $g_{in}(\cdot)$ and $g_{out}(\cdot)$. Using the minimum mean squared error (MMSE) mode, the input function $g_{in}(\cdot)$ is expressed as
$
	g_{in}(\hat{\psi}_{g,r},\tau_{g,r}^\psi,\bm{\Theta})=\mathbb{E}\{t_{g,r}|\mathbf{E},\bm{\Theta},\hat{\psi}_{g,r},\tau_{g,r}^{\psi}\}$,
and the scaled partial derivate of $\tau_{g,r}^\psi g_{in}(\hat{\psi}_{g,r},\tau_{g,r}^\psi,\bm{\Theta})$ w.r.t. $\hat{\psi}_{g,r}$ is
$
	\tau_{g,r}^\psi\frac{\partial}{\partial \hat{\psi}_{g,r}  }g_{in}(\hat{\psi}_{g,r},\tau_{g,r}^\psi,\bm{\Theta}) =\mathbb{V}\{t_{g,r}|\mathbf{E},\bm{\Theta},\hat{\psi}_{g,r},\tau_{g,r}^{\psi}\}$,
where $\mathbb{E}\{\cdot
\}$ and $\mathbb{V}\{\cdot\}$ denote the expectation and variance operations, respectively.

The output scalar function $g_{out}(\cdot)$ is expressed as
$
		g_{out}(\hat{\mu}_{m,r},\tau_{m,r}^\mu,\bm{\Theta})=\frac{\hat{\eta}_{m,r}-\hat{\mu}_{m,r}}{\tau_{m,r}^{\mu}}
		=\frac{e_{m,r}-\hat{\mu}_{m,r}}{\tau_{m,r}^{\mu}+\rho}
$.
Moreover, 
the partial derivate of $g_{out}(\hat{\mu}_{m,r},\tau_{m,r}^{\mu},\bm{\Theta})$ w.r.t. $\hat{\mu}_{m,r}$ is 
$
		\frac{\partial}{\partial\hat{\mu}_{m,r}}g_{out}(\hat{\mu}_{m,r},\tau_{m,r}^{\mu},\bm{\Theta})=\frac{\tau_{m,r}^{\eta}-\tau_{m,r}^{\mu}}{(\tau_{m,r}^{\mu})^2}
		=-\frac{1}{\tau_{m,r}^{\mu}+\rho}$.
Basically, the GAMP flow can be summed up as Algorithm \ref{M-EM-GAMP} lines 5-18 given definitions of $g_{in}(\cdot)$ and $g_{out}(\cdot)$\footnote{The reference \cite{GAMP} provides details on the origin of the GAMP algorithm.}.
\subsection{Hyperparameter Learning via EM}\label{EM}
It is well established that the EM algorithm aims at iteratively maximizing the evidence lower bound (ELBO). We describe briefly the expectation and maximization procedures that derive and optimize ELBO based on the MMV problem.
\subsubsection{Expectation-Step}
From the perspective of maximum likelihood estimate (MLE),
$ \log(\mathbf{E}|\bm{\Theta})$ tends to be maximized w.r.t. $\bm{\Theta}$. This is challenging due to the existence of hidden variables. Thus, EM uses a lower bound w.r.t. ELBO to approach the likelihood, i.e.,
\begin{equation}
	\setlength{\abovedisplayskip}{3.2pt}
	\begin{aligned}
		\log(\mathbf{E}|\bm{\Theta})=&\underbrace{\int_{\bm{\Delta}^{\rm M}}q(\bm{\Delta}^{\rm M})\log\frac{p(\bm{\Delta}^{\rm M},\mathbf{E}|\bm{\Theta})}{q(\bm{\Delta}^{\rm M})} {\rm d}\bm{\Delta}^{\rm M}}_{\rm ELBO} 
		-\underbrace{\int_{\bm{\Delta}^{\rm M}}q(\bm{\Delta}^{\rm M})\log\frac{p(\bm{\Delta}^{\rm M}|\mathbf{E},\bm{\Theta})}{q(\bm{\Delta}^{\rm M})} {\rm d}\bm{\Delta}^{\rm M}}_{\mathrm{KL}(q(\bm{\Delta}^{\rm M})||p(\bm{\bm{\Delta}^{\rm M}}|\mathbf{E},\bm{\Theta}))}\\
		=&\mathrm{ELBO}-\mathrm{KL}(q(\bm{\Delta}^{\rm M})||p(\bm{\Delta}^{\rm M}|\mathbf{E},\bm{\Theta})).
	\end{aligned}
\setlength{\belowdisplayskip}{3.2pt}
\end{equation}
Since the Kullback-Leibler (KL) divergence is nonnegative, in the $o$-th iteration, $\log(\mathbf{E}|\bm{\Theta})$ takes the maximum value when $\mathrm{KL}(q(\bm{\Delta}^{\rm M})||p(\bm{\Delta}^{\rm M}|\mathbf{E},\bm{\Theta}))=0$, i.e., $q(\bm{\Delta}^{\rm M}|\mathbf{E})=p(\bm{\Delta}^{\rm M}|\mathbf{E},\bm{\Theta}^o)$. 

\subsubsection{Maximization-Step}
On the basis of the E-step, $\bm{\Theta}^{o+1}$, where $o+1$ is the current iteration, is solved by 
\begin{equation}
	\setlength{\abovedisplayskip}{3pt}
	\begin{aligned}
		\bm{\Theta}^{o+1}=&\underset{\bm{\Theta}}{\rm arg \ max}\int_{\bm{\Delta}^{\rm M}}p(\bm{\Delta}^{\rm M}|\mathbf{E},\bm{\Theta}^o)\log\frac{p(\bm{\Delta}^{\rm M},\mathbf{E}|\bm{\Theta})}{p(\bm{\Delta}^{\rm M}|\mathbf{E},\bm{\Theta}^o)}{\rm d}\bm{\Delta}^{\rm M}\\
		=&\underset{\bm{\Theta}}{\rm arg \ max} \ \textbf{E}_{\bm{\Delta}^{\rm M}|\mathbf{E},\bm{\Theta}^o}[\log p(\bm{\Delta}^{\rm M},\mathbf{E}|\bm{\Theta})].
	\end{aligned}
	\setlength{\belowdisplayskip}{3pt}
\end{equation}
Due to the difficulty of calculating $p(\bm{\Delta}^{\rm M}|\mathbf{E},\bm{\Theta}^o)$, GAMP is used to attain the approximated  $\hat{p}(\bm{\Delta}^{\rm M}|\mathbf{E},\bm{\Theta}^o)$.
Similar to \cite{EM-GAMP}, the updating rule of $\bm{\Theta}^{o+1}$ is given by
\begin{equation}\label{rho}
	\setlength{\abovedisplayskip}{3.2pt}
	\rho^{o+1}_{r}=\frac{1}{M}\sum_{m=1}^{M}\left(\left|\frac{e_{m,r}-\hat{\mu}_{m,r}}{\tau^\mu_{m,r}/\rho_{r}^{o}+1} \right|^2+\frac{\tau_{m,r}^\mu\rho^o_{r}}{\tau_{m,r}^\mu+\rho^o_{r}}\right),
\setlength{\belowdisplayskip}{3.2pt}
\end{equation}
\begin{equation}\label{nu}
	\setlength{\abovedisplayskip}{3.2pt}
	\nu_{r,l}^{o+1}=\frac{\sum_{g=1}^{G}\varpi_{g,r}^o\upsilon_{g,r,l}^o\varrho_{g,r,l}^o}{\sum_{g=1}^{G}\varpi_{g,r}^o\upsilon^o_{g,r,l}},
	\setlength{\belowdisplayskip}{3.2pt}
\end{equation}
\begin{equation}\label{varsig}
	\setlength{\abovedisplayskip}{3.2pt}
	\varsigma_{r,l}^{o+1}=\frac{\sum_{g=1}^{G}\varpi_{g,r}^o\upsilon_{g,r,l}^o(\vert\nu_{r,l}^o-\varrho_{g,r,l}^o\vert^2+\varkappa^o_{g,r,l})}{\sum_{g=1}^{G}\varpi^o_{g,r}\upsilon^o_{g,r,l}},
	\setlength{\belowdisplayskip}{3.2pt}
\end{equation}
\begin{equation}\label{xi}
	\setlength{\abovedisplayskip}{3.2pt}
	\omega^{o+1}_{r,l}=\frac{\sum_{g=1}^{G}\varpi_{g,r}^o\upsilon^o_{g,r,l}}{\sum_{g=1}^{G}\varpi^o_{g,r}}.
	\setlength{\belowdisplayskip}{3.2pt}
\end{equation}
Particularly, the sparsity factor is determined by the average of $R$ measurement vectors, i.e.,
\begin{equation}\label{kappa}
	\setlength{\abovedisplayskip}{3.2pt}
	\kappa^{o+1}_{g,r}=\frac{1}{R}\sum_{r=1}^{R}\frac{1}{1+\left(\frac{\sum_{l=1}^{L}\kappa_{g,r}^o \omega_{r,l}^o\mathcal{CN}(\hat{\psi}_{g,r}|\nu^o_l,\varsigma^o_l+\tau_{g,r}^\psi)}{(1-\kappa_{g,r}^o)\mathcal{CN}(0|\hat{\psi}_{g},\tau_{g,r}^\psi)}\right)^{-1}}.
	\setlength{\belowdisplayskip}{3.2pt}
\end{equation}

In summary, the EM procedure can be found in Algorithm \ref{M-EM-GAMP} lines 21-24.
\subsection{LoS-Aided Off-Grid Refinement}\label{OFF-GRID}
Note that the sensing matrix $\mathbf{A}_{\rm F/H}$ in Eqn. (\ref{DBS2}) is comprised of two matrices. Hence, we suppose that $\bm{\Phi}=\mathbf{C}\widetilde{\mathbf{A}}$, where $\mathbf{C}\in\mathbb{C}^{M\times N}$ is the measurement matrix (corresponding to the transmit/receive beams), and $\widetilde{\mathbf{A}}\in\mathbb{C}^{N\times G}$ is the partial dictionary regarding Eqn. (\ref{UPAs}) for virtual channel representation, in which each column represents a recovered atom. 

Assuming that Eqn. (\ref{E2}) is solved by the previous recovery procedure, and that the angle set $\{\theta_b^{\rm ele/azi}\}_{b=1}^P$ of $\widetilde{\mathbf{A}}$
is estimated,
the following perturbation-based off-grid problem is formed:
\begin{equation}\label{E3}
	\setlength{\abovedisplayskip}{3.2pt}
		\underset{\bm{\mathcal{P}}_{\theta^{\rm ele/azi}}, \widetilde{\mathbf{\bm{\Delta}}}^{\rm M}}{\rm arg \ min}  \left\|\mathbf{E}-\mathbf{C}\sum_{b=1}^{P}\mathbf{a}\left(  \cos(\theta ^{\rm ele}_{b})(1+\mathcal{P}_{\theta ^{\rm ele},b}), 
	\sin(\theta_b^{\rm ele})\sin(\theta_b^{\rm azi} )(1+\mathcal{P}_{\theta^{\rm azi},b})\right) \widetilde{\mathbf{t}}_b^T\right\|_F^2,
	\setlength{\belowdisplayskip}{3.2pt}
\end{equation}
where $\frac{2d}{\lambda}=1$ is set, $\bm{\mathcal{P}}_{\theta^{\rm ele}}=[\mathcal{P}_{\theta^{\rm ele},1},\cdots,\mathcal{P}_{\theta^{\rm ele},P}]$ and $\bm{\mathcal{P}}_{\theta^{\rm azi}}=[\mathcal{P}_{\theta^{\rm azi},1},\cdots,\mathcal{P}_{\theta^{\rm azi},P}]$ are the perturbations, and $\widetilde{\mathbf{\bm{\Delta}}}^{\rm M}=[\widetilde{\mathbf{t}}_1,\cdots,\widetilde{\mathbf{t}}_P]^T\in\mathbb{C}^{P\times R}$ is the estimated sparse signal. 

Note that the LoS path  
does not need to be refined due to LoS-aided dictionaries used. Thus, the off-grid procedure is simplified. For clarity, the path index $b=1$ is assumed to be the LoS path. 
Denoting $\cos(\theta^{\rm ele})$ and $\sin(\theta^{\rm ele})\sin(\theta^{\rm azi})$ by $\widetilde{\theta}^{\rm ele}$ and $\widetilde{\theta}^{\rm azi}$, respectively. Based on the first-order Taylor expansion,
 we have
\begin{equation}
	\setlength{\abovedisplayskip}{3.2pt}
	\underset{\bm{\mathcal{P}}_{\theta^{\rm ele}},\bm{\mathcal{P}}_{\theta^{\rm azi}},\widetilde{\mathbf{\bm{\Delta}}}^{\rm M}}{\rm arg \ min} \ \left\| \mathbf{R}-\mathbf{C}
	\sum_{b=2}^{P}\left(\mathcal{P}_{\theta ^{\rm ele},b}\widetilde{\theta}^{\rm ele}_b\frac{\partial\mathbf{a}\left(\widetilde{\theta}_b^{\rm ele},\widetilde{\theta}_b^{\rm azi}\right)}{\partial\widetilde{\theta}^{\rm ele}_b}\widetilde{\mathbf{t}}_b^T \right.\right. 
 \left.\left. + \mathcal{P}_{\theta^{\rm azi},b}	\widetilde{\theta}_b^{\rm azi}\frac{\partial\mathbf{a}\left(\widetilde{\theta}_b^{\rm ele},\widetilde{\theta}_b^{\rm azi}\right)}{\partial\widetilde{\theta}_b^{\rm azi}}\widetilde{\mathbf{t}}_b^T\right)\right\|_F^2,
 \setlength{\belowdisplayskip}{3.2pt}
\end{equation}
where $
\mathbf{R}=\mathbf{E}-\mathbf{C}\mathbf{a}\left(\widetilde{\theta}_1^{\rm ele},\widetilde{\theta}_1^{\rm azi}\right)\widetilde{\mathbf{t}}_1^T-\mathbf{C}\sum_{b=2}^{P}\mathbf{a}\left(\widetilde{\theta}_b^{\rm ele},\widetilde{\theta}_b^{\rm azi}\right)\widetilde{\mathbf{t}}_b^T$
 is the residual. For this coupled optimization, the alternating iteration algorithm is used. We first present the following theorem w.r.t. the matrix-based least squares (MLS) algorithm.

\textbf{Theorem 2:} Postulate that $\mathbf{Y}\in\mathbb{C}^{M\times R}$, $\mathbf{A}_b\in\mathbb{C}^{M\times R}, b\in\{1,2,\cdots,P\} \ (P\leq MR)$ and
$\mathbf{w}=[w_1,w_2\cdots,w_P]\in\mathbb{C}^{P\times 1}$, the closed-form solution w.r.t. $\mathbf{w}$ of 
$
\underset{\mathbf{w}}{\rm arg \ min} \ \left\| \mathbf{Y}-\sum_{b=1}^Pw_b\mathbf{A}_b\right\|_F^2
$
is 
$
\mathbf{w}=(\mathbf{G}^H\mathbf{G})^{-1}\mathbf{G}^H{\rm vec}(\mathbf{Y})
$
with $\mathbf{G}=[{\rm vec}(\mathbf{A}_1),{\rm vec}(\mathbf{A}_2),\cdots,{\rm vec}(\mathbf{A}_P)]\in\mathbb{C}^{MR\times P}$.

\textbf{Proof:} See Appendix \ref{appendixB}.

\begin{algorithm}[!t] 
	\caption{Proposed Two-Timescale Estimation Procedure } 
	\label{overall}      
	\begin{algorithmic}[1] 
		\footnotesize{
			\REQUIRE { LoS path's angle set $\{\overline{\vartheta}^{\rm ele/azi}_{i},\overline{\varphi}^{\rm ele/azi}_{i},\overline{\theta}^{\rm ele/azi},\overline{\phi}^{\rm ele/azi}|i=1,2\}$, received signals via uplink training.
			}

			\ENSURE {Channels $\mathbf{F}_i$, $\mathbf{D}$, $\mathbf{h}_{i,u}$}

			\STATE{\textbf{\%\% Large-Timescale Estimation}}
			\STATE{\emph{Estimation of the BS-RIS channel $\mathbf{F}_i, i\in\{1,2\}$}:}
			\STATE{  Solve Eqn. (\ref{RIS_yu}) at the RIS with the off-grid EM-GAMP algorithm, and feed back to the BS;}
			\STATE{ Construct $\{\overline{\vartheta}^{\rm ele/azi}_i,\overline{\varphi}^{\rm ele/azi}_i\}$-aided dictionaries as Eqn. (\ref{Ax}); 
			}
			\STATE{Decouple Eqn. (\ref{Yddot}) into two parallel problems according to Eqn. (\ref{E});}
			\STATE{Invoke Algorithm \ref{M-EM-GAMP} to solve the above problems to attain the BS-RIS channel' AoAs, AoDs and path gains according to Eqn. (\ref{DD}).}

			\STATE{\emph{Estimation of the RIS 1-RIS 2 channel $\mathbf{D}$:} 
			}
			\STATE{Remove the signals regarding the BS-RIS 1/2-user channel to get Eqn. (\ref{DBS1});}
			\STATE{ Construct $\{\overline{\theta}^{\rm ele/azi},\overline{\phi}^{\rm ele/azi}\}$-aided dictionaries as Eqn. (\ref{Ax}); 
			}
			\STATE{Decouple Eqn. (\ref{Yddot}) into two parallel problems according to Eqn. (\ref{E});}
			\STATE{Invoke Algorithm \ref{M-EM-GAMP} to solve the above problems to attain the RIS 1-RIS 2 channel' AoAs, AoDs and path gains according to Eqn. (\ref{DD}).}
			\STATE{\textbf{\%\% Small-Timescale Estimation}}
			\STATE{Solve Eqn. (\ref{T2}) using the off-grid EM-GAMP algorithm. Note that the LoS-aided method is not applicable here since we don't assume we know the location of all users. }

		}
		
	\end{algorithmic}
\end{algorithm}

Based the above theorem, the updating rules w.r.t. $\widetilde{\theta}^{\rm ele}_b$ and $\widetilde{\theta}^{\rm azi}_b$ $(\forall b,b\neq 1)$   are given by
\begin{equation}\label{thetat1}
	\setlength{\abovedisplayskip}{3.2pt}
\widetilde{\theta}_b^{{\rm ele},t+1}=
 	\widetilde{\theta}_b^{{\rm ele},t}+\textbf{Re}\{[(\widetilde{\mathbf{G}}_{\theta^{{\rm ele},t}}^H\widetilde{\mathbf{G}}_{\theta^{{\rm ele},t}})^{-1}\widetilde{\mathbf{G}}_{\theta^{{\rm ele},t}}^H{\rm vec}(\mathbf{R}^t)]_b\}\widetilde{\theta}_b^{{\rm ele},t},
 	\setlength{\belowdisplayskip}{3.2pt}
\end{equation}
\begin{equation}\label{theta2}
	\setlength{\abovedisplayskip}{3.2pt}
 \widetilde{\theta}_b^{{\rm azi},t+1}=
 \widetilde{\theta}_b^{{\rm azi},t}+\textbf{Re}\{[(\widetilde{\mathbf{G}}_{\theta^{{\rm azi},t}}^H\widetilde{\mathbf{G}}_{\theta^{{\rm azi},t}})^{-1}\widetilde{\mathbf{G}}_{\theta^{{\rm azi},t}}^H{\rm vec}(\mathbf{R}^t)]_b\}\widetilde{\theta}_b^{{\rm azi},t},
 \setlength{\belowdisplayskip}{3.2pt}
\end{equation}
where $\widetilde{\mathbf{G}}_{\theta^{{\rm ele},t}}=[{\rm vec}({\mathbf{G}}_{\theta^{{\rm ele},t},2}),\cdots{\rm vec}({\mathbf{G}}_{\theta^{{\rm ele},t},P})]$ with $
{\mathbf{G}}_{\theta^{{\rm ele},t},b}=\widetilde{\theta}_b^{{\rm ele},t}\mathbf{C}\frac{\partial\mathbf{a}\left(\widetilde{\theta}_b^{{\rm ele},t},\widetilde{\theta}_b^{{\rm azi},t}\right)}{\partial\widetilde{\theta}^{{\rm ele},t}_b}\widetilde{\mathbf{t}}_b^{t,T}
$
and $\widetilde{\mathbf{G}}_{\theta^{{\rm azi},t}}=[{\rm vec}({\mathbf{G}}_{\theta^{{\rm azi},t},2}),\cdots{\rm vec}({\mathbf{G}}_{\theta^{{\rm azi},t},P})]$ with
$
	{\mathbf{G}}_{\theta^{{\rm azi},t},b}=\widetilde{\theta}_b^{{\rm azi},t}\mathbf{C}\frac{\partial\mathbf{a}\left(\widetilde{\theta}_b^{{\rm ele},t},\widetilde{\theta}_b^{{\rm azi},t}\right)}{\partial\widetilde{\theta}^{{\rm azi},t}_b}\widetilde{\mathbf{t}}_b^{t,T}
$.
In particular, according to Eqn. (\ref{UPAs}), $\frac{\partial\mathbf{a}\left(\widetilde{\theta}_b^{{\rm ele},t},\widetilde{\theta}_b^{{\rm azi},t}\right)}{\partial\widetilde{\theta}^{{\rm ele},t}_b}$ and $\frac{\partial\mathbf{a}\left(\widetilde{\theta}_b^{{\rm ele},t},\widetilde{\theta}_b^{{\rm azi},t}\right)}{\partial\widetilde{\theta}^{{\rm azi},t}_b}$ can be calculated by
\begin{equation}
	\setlength{\abovedisplayskip}{3.2pt}
	\frac{\partial\mathbf{a}(\widetilde{\theta}_b^{{\rm ele},t},\widetilde{\theta}_b^{{\rm azi},t})}{\partial\widetilde{\theta}^{{\rm ele},t}_b}=\sqrt{\frac{1}{N_yN_z}}([1,\cdots,e^{j\pi(N_z-1)\widetilde{\theta}_b^{{\rm ele},t}}]^T  \odot \left[ 1,\cdots,j\pi (N_z-1)\right]^T
	)
		\otimes[1,\cdots,e^{j\pi(N_y-1)\widetilde{\theta}_b^{{\rm azi},t}}]^T,
		\setlength{\belowdisplayskip}{3.2pt}
\end{equation} 
\begin{equation}
	\setlength{\abovedisplayskip}{3.2pt}
		\frac{\partial\mathbf{a}(\widetilde{\theta}_b^{{\rm ele},t},\widetilde{\theta}_b^{{\rm azi},t})}{\partial\widetilde{\theta}^{{\rm azi},t}_b}=\sqrt{\frac{1}{N_yN_z}}[1,\cdots,e^{j\pi(N_z-1)\widetilde{\theta}_b^{{\rm ele},t}}]^T   \otimes ( [ 1,\cdots,j\pi (N_y-1)]^T
		\odot[1,\cdots,e^{j\pi(N_y-1)\widetilde{\theta}_b^{{\rm azi},t}}]^T).
		\setlength{\belowdisplayskip}{3.2pt}
\end{equation}

After $\widetilde{\theta}^{\rm ele}$ and $\widetilde{\theta}^{\rm azi}$ are updated, the partial dictionary matrix w.r.t. angles is also updated as $\widetilde{\mathbf{A}}^{t+1}$. Then the coefficient matrix and the residual are renewed by 
\begin{equation}\label{DLS}
	\setlength{\abovedisplayskip}{3.2pt}
\widetilde{\bm{\Delta}}^{{\rm M},t+1}= (\widetilde{\mathbf{A}}^{t+1,H}\mathbf{C}^H\mathbf{C}\widetilde{\mathbf{A}}^{t+1})^{-1}\widetilde{\mathbf{A}}^{t+1,H}\mathbf{C}^H\widetilde{\mathbf{E}},
\setlength{\belowdisplayskip}{3.2pt}
\end{equation}
\begin{equation}\label{DRES}
	\setlength{\abovedisplayskip}{3.2pt}
\mathbf{R}^{t+1}=\widetilde{\mathbf{E}}-\mathbf{C}\widetilde{\mathbf{A}}^{t+1}\widetilde{\bm{\Delta}}^{{\rm M},t+1},
\setlength{\belowdisplayskip}{3.2pt}
\end{equation}
where $\widetilde{\mathbf{E}}=\mathbf{E}-\mathbf{C}\mathbf{a}\left(\widetilde{\theta}_1^{\rm ele},\widetilde{\theta}_1^{\rm azi}\right)\widetilde{\mathbf{t}}_1^T$ is the signal after the LoS path is removed.
Then the off-grid procedure performs $\mathcal{T}$ iterations. For a stable solution, we set an iteration termination condition that stops the off-grid procedure when $\mathbf{R}^{t+1}>\mathbf{R}^t$.
Finally, the estimated sparse signal $\widehat{\bm{\Delta}}^{\rm M}$ can be obtained by filling zero rows into $\widetilde{\bm{\Delta}}^{{\rm M},\mathcal{T}}$.
In summary, the whole two-timescale channel estimation process is shown in Algorithm. \ref{overall}.

\section{Simulation Results}\label{S5}
This section supports simulations to determine the effectiveness of our proposed strategies. To begin, we define the normalized mean square error (NMSE) as $\mathbb{E}\{\frac{\Vert\widehat{\mathbf{D}}-\mathbf{D}\Vert_F^2}{\Vert \mathbf{D}\Vert_F^2}\}$, where $\widehat{\mathbf{D}}$ is the estimated RIS 1-RIS 2 channel. The NMSE definition of other channels is similar to that of the above.
The system setup considered for our simulations consists of $J=36$, $L_1=L_2=64$, $U=4$ and system frequency $f=28$ GHz with bandwidth $B=100$ MHz . The BS and RISs are all assumed to be equipped with uniform square arrays. The noise figure $(NF)$ is set to 9. Hence, the noise power is given by $\sigma_n^2=-174 \ {\rm dBm/Hz}+10{\rm log_{10}}(B)+NF$.
 For channel parameters, we employ the $28$ GHz mmWave channel setting in \cite{channel}. Following Table. I in \cite{channel}, we assume that that path gains follow $\mathcal{CN}(0,\aleph10^{-0.1{\rm PL}})$, where $\aleph=K_1^{1.8}10^{0.1K_2}$ with $K_1\sim \mathcal{U}(0,1)$ and $K_2\sim \mathcal{CN}(0,16)$. Besides, ${\rm PL}=a_1+10a_2{\rm log}_{10}(d^\prime)+\mathcal{N}(0,\sigma) [\rm{dB}]$ with $d^\prime$ denoting the distance in meters, where $a_1=72$, $a_2=2.92$ and $\sigma=8.7$ for NLoS paths, and $a_1=61.4$, $a_2=2$ and $\sigma=5.8$ for LoS paths. For the deployment of the BS and RISs, we assume the three-dimensional Cartesian coordinates of the BS, RIS 1 and RIS 2 are $(0,0,5)$, $(10\sqrt{2},10\sqrt{2},6)$, $(10\sqrt{2}+100,10\sqrt{2},6)$, and the users are uniformly distributed  around RIS 2 at a distance of $1$ to $30$ meters. Moreover, we assume $P_{f_i}=P_d=P_{h,i,u}=3$ for $\forall i,u$. The LoS paths of BS-RISs and RIS 1-RIS 2 can be caculated by their locations. The LoS paths of users' channels and NLoS paths of all channels are uniformly distributed in elevation and azimuth coverage. All users' uplink power $\sigma_p^2$ is identical.
 Lastly, the size of dictionaries is set as $G_t=J$ and $G_r=L$.
 \begin{figure*}
 	\centering
 	\subfigure[BS-RIS 1 Channel, $\sigma_p^2$ = 30 dBm]{
 		\includegraphics[width=0.305\textwidth]{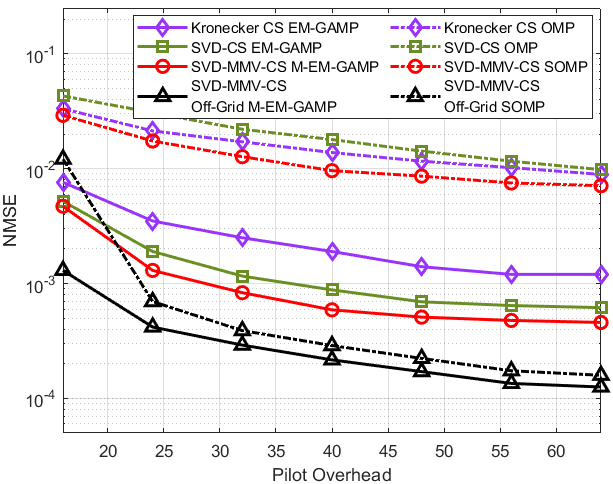}
 	}
 	\subfigure[BS-RIS 2 Channel, $\sigma_p^2$ = 30 dBm]{
 		\includegraphics[width=0.305\textwidth]{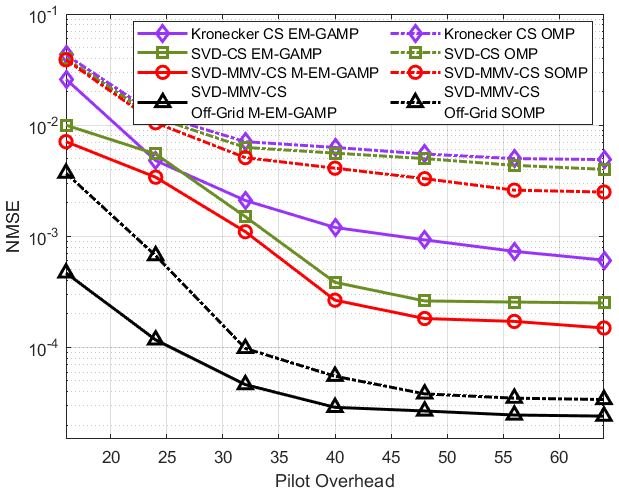}
 	}
 	\subfigure[RIS 1-RIS 2 Channel, $\sigma_p^2$ = 30 dBm]{
 		\includegraphics[width=0.305\textwidth]{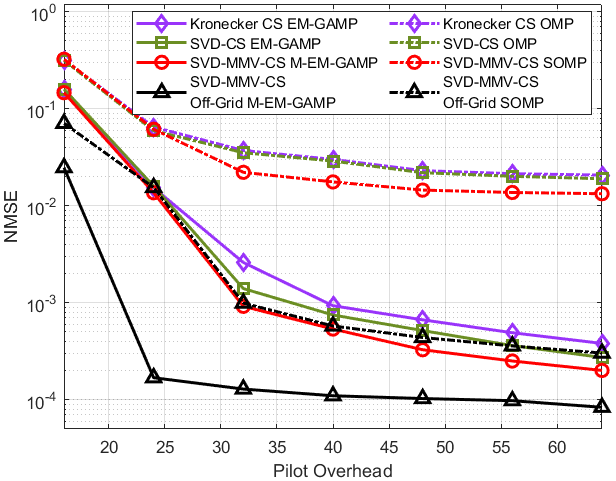}
 	}
 	\caption{NMSE performance of different schemes versus pilot overhead at $\sigma_p^2$ = 30 dBm.}
 	\label{L30}
 \end{figure*}

\subsection{Evaluation of Large-Timescale Channel Estimation}\label{Simulation-largescale}
Here, we will highlight the excellence of our proposed large-timescale channel estimation method. To this end, we will evaluate our proposed  low-complexity Bayesian framework and the MAE method for large-timescale estimation. 
Moreover, we use the greedy methods, orthogonal match pursuit (OMP)/simultaneous orthogonal match pursuit (SOMP), as the algorithm benchmark. 

\begin{figure*}
	\centering
	\subfigure[BS-RIS 1 Channel, $\sigma_p^2$ = 15 dBm]{
		\includegraphics[width=0.305\textwidth]{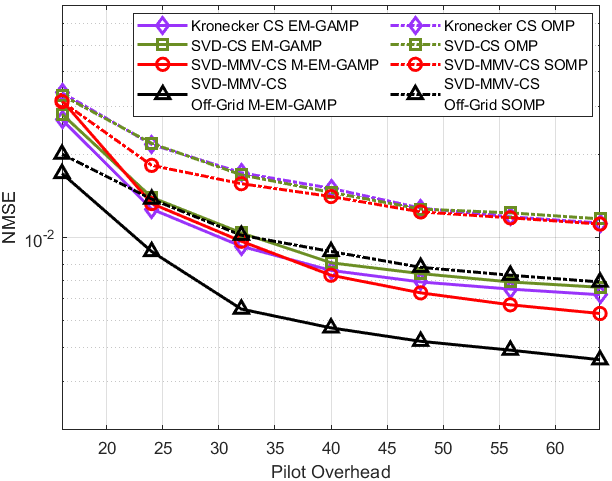}
	}
	\subfigure[BS-RIS 2 Channel, $\sigma_p^2$ = 15 dBm]{
		\includegraphics[width=0.305\textwidth]{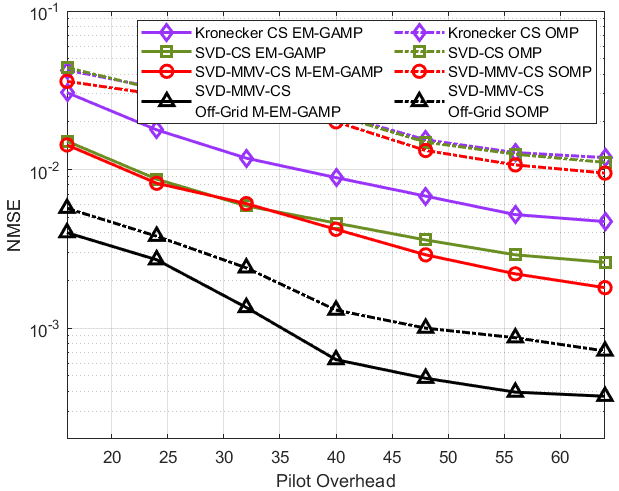}
	}
	\subfigure[RIS 1-RIS 2 Channel, $\sigma_p^2$ = 15 dBm]{
		\includegraphics[width=0.305\textwidth]{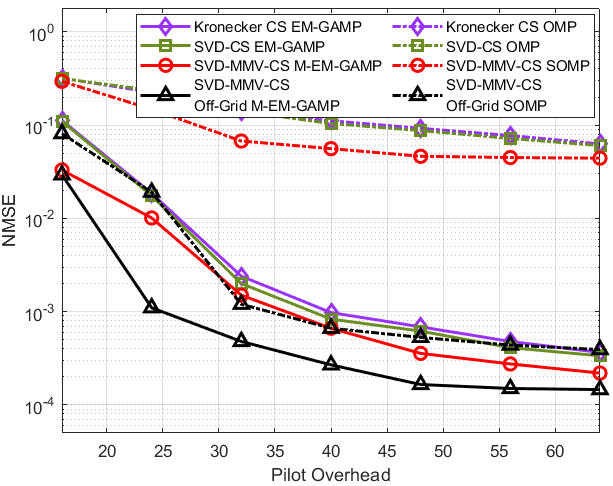}
	}
	\caption{NMSE performance of different schemes versus pilot overhead at $\sigma_p^2$ = 15 dBm.}
	\label{L15}
\end{figure*}
\begin{figure}[htbp]
	\centering
	\subfigure[NMSE performance at $\sigma_p^2$ = 30 dBm]{
		\includegraphics[width=2.6in]{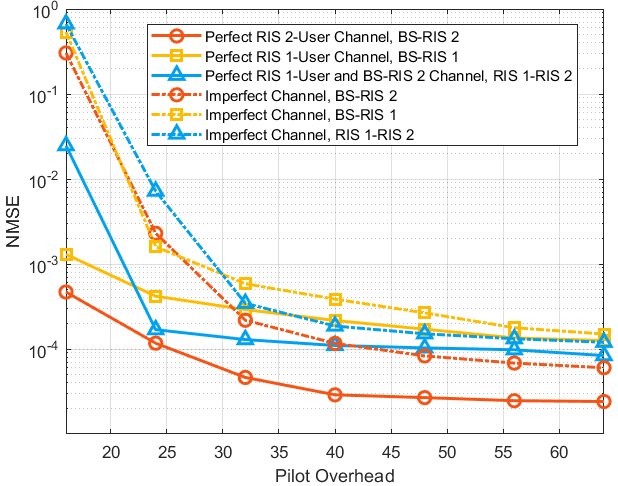}
	}
	\quad    
	\subfigure[NMSE performance at $\sigma_p^2$ = 15 dBm]{
		\includegraphics[width=2.6in]{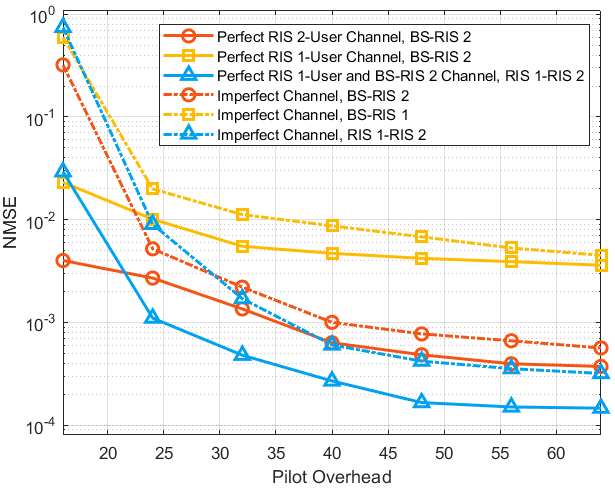}
	}
	\caption{NMSE performance of perfect/imperfect channels-based large-timescale channel estimation versus pilot overhead at $\sigma_p^2$ = 15/30 dBm.}
	\label{largeIMP}
\end{figure}
\subsubsection{CS Framework Comparison}\label{Simulation-complexity}
We first evaluate the effectiveness of our proposed CS frameworks, i.e., SVD-MMV-CS and off-grid SVD-MMV-CS frameworks. 
For a good comparison, the Kronecker CS and SVD-CS frameworks are employed as benchmarks. Note that Eqn. (\ref{Yddot}) and (\ref{DBS1}) depend on the RIS-user channels, i.e., $\mathbf{h}_{i,u}, \forall i,u$, here we assume they are accurately estimated for CS framework comparison. Certainly, the impact of these channels' estimation accuracy will be explored later. Firstly, Fig. \ref{L30} exhibits the NMSE performance of different schemes regarding the estimation of the BS-RIS 1 channel, BS-RIS 2 channel and RIS 1-RIS 2 channel, where uplink power $\sigma_p^2$ = 30 dBm and the pilot overhead $Q_1=Q_2$ ranges from 16 to 64. Particularly, the pilot overhead $Q_3=N_X\times N_Y$ for RIS 1-RIS 2 channel estimation follows that $N_X=N_Y$ ranges from 16 to 64. It can be observed that the proposed SVD-MMV-CS framework outperforms the SVD-CS and Kronecker CS frameworks in terms of the NMSE. Besides, the NMSE performance of on-grid EM-GAMP-based methods is much better than that of on-grid OMP-based methods. Furthermore, the off-grid technique based on the proposed framework can achieve a significant performance improvement (please see the red and black curves in Fig. \ref{L30}). Notably, the NMSE of Fig. \ref{L30}. (a) is higher than that of Fig. \ref{L30}. (b). This is because the path loss of the BS-RIS 2-user channel is generally smaller than that of the BS-RIS 1-user channel.
 To evaluate the impact of the uplink power on the NMSE performance of different schemes, we set $\sigma_p^2=15$ dBm to repeat the above experiments in Fig. \ref{L15}. It can be shown that all schemes' NMSEs become larger. Despite this, the off-grid M-EM-GAMP algorithm based on the SVD-MMV-CS framework has a considerable performance at a low pilot overhead value.

\subsubsection{Estimation Strategy Evaluation}

Note that the uplink training of the BS-RIS $i$ channel is reliant on the previous step of estimating $\mathbf{h}_{i,u}, \forall i,u$. Besides, the estimation of $\mathbf{D}$ depends on $\mathbf{h}_{1,u}$ and $\mathbf{F}_2$
according to Eqn. (\ref{DBS2}). Hence, we will explore the impact of $\mathbf{h}_{i,u}$ and $\mathbf{F}_2$ on the estimation of the BS-RIS 1, BS-RIS 2 and RIS 1-RIS 2 channels. 
Toward this end, we compare the NMSE performance of different schemes under the perfect channel assumption and the imperfect channel assumption, respectively. The imperfect channel assumption means that the channel required to estimate the large-timescale channel is obtained by the specified method. For example, when estimating $\mathbf{F}_i$, the required channel $\mathbf{h}_{i,u}$ is obtained by using the off-grid EM-GAMP method to solve Eqn. (\ref{RIS_yu}) instead of being perfectly known. It is noteworthy that when estimating $\mathbf{D}$, the required channel $\mathbf{F}_2$ is obtained under the imperfect channel assumption. As shown in Fig. \ref{largeIMP} with uplink power $\sigma_p^2=30$ and $15$
dBm, the NMSE performance under the perfect channel assumption is much better than that under the imperfect channel assumption when the pilot overhead is small. As the pilot overhead increases, the NMSE gap between the two assumptions becomes narrow, indicating that the impact of the required channels on large-timescale channel estimation becomes small as the pilot overhead becomes large. This is because the estimation accuracy of the required channel increases with the increase of the pilot overhead.

\begin{figure}
	\centering
	\subfigure[NMSE performance of RIS 1-user channel estimation]{
		\includegraphics[width=2.6in]{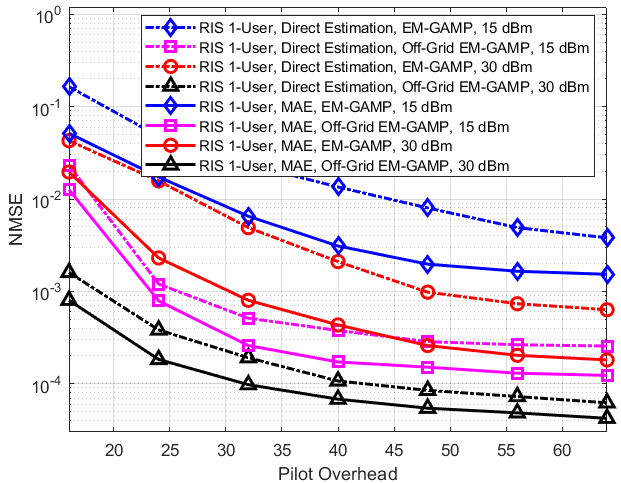}
	}
	\quad    
	\subfigure[NMSE performance of RIS 2-user channel estimation]{
		\includegraphics[width=2.6in]{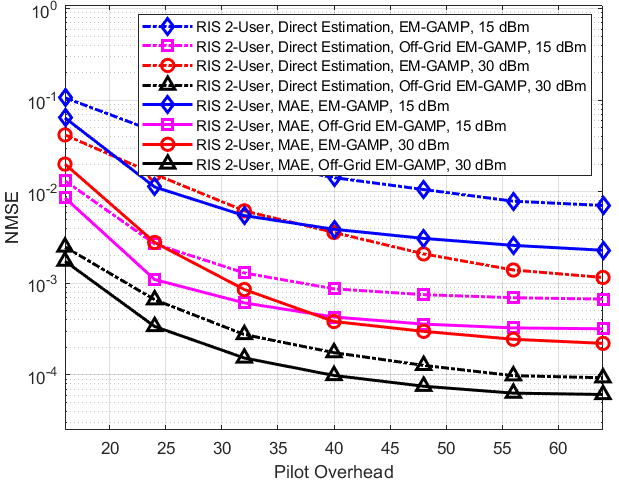}
	}
	\caption{NMSE performance of direct estimation- and MAE-based small-timescale channel estimation versus pilot overhead at $\sigma_p^2$ = 15/30 dBm.}
	\label{T2MAE}
\end{figure}

\subsection{Evaluation of Small-Timescale Channel Estimation}\label{Simulation-smallscale}

\begin{figure}
 	\centering
 	\includegraphics[width = 0.415\textwidth]{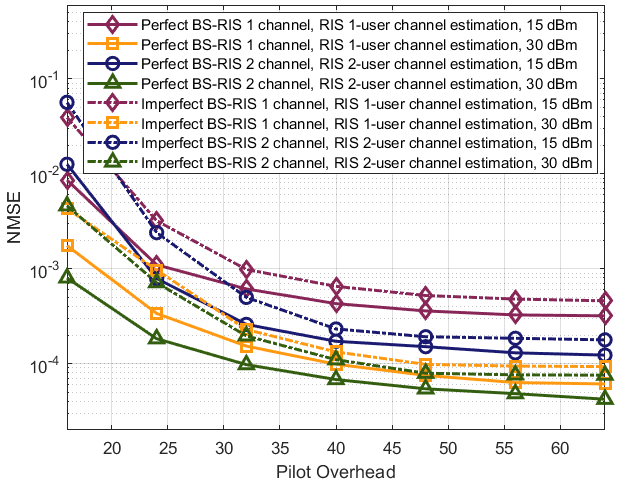}
 	\caption{NMSEs of small-timescale channel estimation with perfect/imperfect channel.}
 	\label{smallIMP}
 \end{figure}

To show the effectiveness of the MAE method for small-timescale channel estimation, we use the method of directly estimating the RIS-user channel at the RIS using the receive RF chain as the benchmark, which is the first step of the estimation of large-timescale channels.
Seeing Fig. \ref{T2MAE}, in which the pilot overhead ranges from 16 to 64, and $\sigma_p^2$ is set to 15 and 30 dBm, it can be concluded that the NMSE performance of the MAE-based methods is better than the direct estimation-based methods due to the former's higher dimensional sensing matrix. When using the direct estimation strategy, the NMSE performance of Fig. \ref{T2MAE}. (a) is slightly better than that of Fig. \ref{T2MAE}. (b) since users are close to RIS 1 than to RIS 2. Furthermore, Fig. \ref{T2MAE} indicates that our proposed MAE strategy used in the small-timescale stage can reduce the pilot overhead when estimating the RIS-user channel. That is, when achieving an NMSE performance index, the MAE strategy requires less pilot overhead than the direct estimation strategy, $\bar{Q}_i<Q_i, \ i\in\{1,2\}$.
 Additionally, we evaluate the impact of the large-timescale estimation performance on the MAE-based small-timescale estimation strategy.
 In Fig. \ref{smallIMP}, the NMSE performance of different schemes under the perfect and imperfect channel assumptions is shown with different unplink power. Here, the imperfect channel assumption denotes the large-scale channel required for RIS-user channel estimation is obtained by the performance-optimal SVD-MMV-CS-based off-grid M-EM-GAMP algorithm.
 It can be observed that the perfect channel assumption-based NMSE performance is better than the imperfect channel assumption-based NMSE performance. As the pilot overhead increases, the NMSE performance of the two assumptions-based schemes is gradually approaching. 

\subsection{Complexity Analysis}
Table. \ref{table1} provides a comparison of various schemes regarding pilot overhead and computational complexity, where the recovery algorithm options \ding{172} and \ding{173} denote the the variant algorithms of OMP and EM-GAMP, respectively.  Notably, all schemes are conducted by two algorithms, i.e., \ding{172} and \ding{173}. We will discuss the computing cost of OMP-based and EM-GAMP-based methods' successively.
First, we denote the iteration number for the two algorithms and the off-grid method by $I_{ 1}$, $I_{ 2}$ (both the GAMP and EM loops are considered) and $I_3$, respectively.

\subsubsection{Large-Timescale}

We first consider the pilot overhead of large-timescale estimation. Due to the fact that the large-timescale channels are estimated only once in a timescale period, the average pilot overhead is $\frac{S_2}{S_1}(Q_1+Q_2+Q_3)$, with $S_1$ and $S_2$ denoting the large-timescale and small-timescale channel coherence time, respectively.

Since the computing cost of estimating $\mathbf{D}$ dominates the whole large-scale channel estimation procedure, the computational complexity of estimating $\mathbf{D}$ is analyzed in detail. Recalling Eqn. (\ref{DBS2}), the kronecker CS framework will process the large-scale sensing matrix. It is well established that the atom matching step dominates the complexity of OMP, which will incur a per-iteration complexity of $N_XN_YUJ G_r^2$ by calculating the projection of $\mathbf{A}_{\rm H}^T\otimes\mathbf{A}_{\rm F}$ onto $\bar{\mathbf{y}}^{\rm BS}$. Thus, its whole complexity is $\mathcal{O}(I_{1}N_XN_YUJ G_r^2)$. On the contrary, the SVD-CS framework deals with two parallel subproblems with small-scale sensing matrices. Thus, this two subproblems will need to calculate two simple projections with $N_XUG_rR$ and $N_YJG_rR$, where $R$ is the rank of the received signal and is set to $3$\footnote{Indeed, $R$ exactly equals to the number of paths, i.e., the channel's rank, with the noiseless received signal. Although considering the noise effect, we set $R$ to the number of paths, i.e., $R = 3$, for simplicity.}. On the other hand, $N_X$ and $N_Y$ are the number of measurements of the two RISs, they should be approximately equal. Generally, the number of access users is not greater than that of BS's antennas, i.e., $U\leq J$. Hence, $N_XU$ and $N_YJ$ are on the same order of magnitude.
 Thus, the whole complexity can be expressed as $\mathcal{O}(I_1N_YG_rJ)$. Furthermore, our proposed off-grid SVD-MMV-CS incurs the same complexity as SVD-CS. The difference between them is when processing all measurement vectors' atom projections, SVD-CS can calculate them in parallel, while SVD-MMV-CS requires all measurement vectors to calculate them jointly. Next, we will discuss the complexity of our proposed off-grid method based on the SVD-MMV-CS framework. 
  Let us consider Eqn. (\ref{thetat1}) for the off-grid solutions of the two subproblems, it will produce a complexity of $N_YJR(P-1)^2+(P-1)^3$ and $N_XUR(P-1)^2+(P-1)^3$, respectively. $(P-1)^3$ produced by the inverse operation can be ignored because the small estimated number of paths $P-1$\footnote{Due to the known LoS path, the number of paths requiring off-grid refinement is reduced to $P-1$.}. 
  Considering Eqn. (\ref{DLS}) yields a different complexity of $N_YJL(P-1)$ or $N_XUL(P-1)$. Note that $(P-1)R$ is generally slighter than $L$ since both $P-1$ and $R$ are related to the low rank of the channel. In addition, according to the relation between $N_XU$ and $N_YJ$ discussed before, the whole complexity of the proposed off-grid method is summarized as $\mathcal{O}(I_3N_YJL(P-1))$. Similar to above manner, the OMP-based complexity of the other estimation of large-scale channels can be derived. Next, we will provide the complexity of large-scale channel estimation regarding the algorithm option \ding{173}. According to \cite{EM-GAMP}, the per-iteration complexity of EM-GAMP scales as that of OMP. The difference is that EM-GAMP requires more iterations than OMP, i.e., generally $I_2\gg I_1$. Due to the small number of multiple measurement vectors $R=3$, M-EM-GAMP has a comparable complexity with EM-GAMP. Therefore,
  the complexity derivation of all schemes with respect to \ding{173} is similar to that with respect to \ding{172}.

\begin{table*}\footnotesize
	\renewcommand\arraystretch{1.05} 
	\caption{A Comparison of Pilot Overhead and Computational Complexity of Various Schemes}
	\label{table1}
	\centering
	\begin{tabular}{ |c|c|c|c|c|}
		\hline
		\tabincell{c}{Two-Timescale\\ Estimation}	&\tabincell{c}{Average \\ Pilot Overhead}&	CS Framework &\tabincell{c}{Recovery\\ Algorithm }
		& \tabincell{c}{ Computational\\ Complexity }\\ \hline\hline
		
		\multirow{8}{*}{\tabincell{c}{Large-Timescale \\ Estimation}}		& \multirow{8}{*}{\tabincell{c}{$\frac{S_2}{S_1}(Q_1+Q_2+Q_3)$}}& \multirow{2}{*}{Kronecker CS} &\ding{172}&$\mathcal{O}(I_{1}N_XN_YUJ G_r^2)$
		\\ \cline{4-5}
		& & &\ding{173}&$\mathcal{O}(I_{2}N_XN_YUJ G_r^2)$\\
		\cline{3-5}
		& &\multirow{2}{*}{SVD-CS}&\ding{172}&$\mathcal{O}(I_1N_YG_rJ)$\\  \cline{4-5}
		& & &\ding{173}&$\mathcal{O}(I_2N_YG_rJ)$ \\
		\cline{3-5}
		& & \multirow{2}{*}{\tabincell{c}{Proposed On-Grid \\ SVD-MMV-CS}}&\ding{172}&$\mathcal{O}(I_1N_YG_rJ)$ \\
		\cline{4-5}
		& & &\ding{173}&$\mathcal{O}(I_2N_YG_rJ)$ \\
		\cline{3-5}
		& & \multirow{2}{*}{\tabincell{c}{ Proposed Off-Grid \\ SVD-MMV-CS}}&\ding{172}&\tabincell{c}{$\mathcal{O}(I_1N_YG_rJ+I_3N_YJL(P-1))$ } \\
		\cline{4-5}
		& & &\ding{173} &\tabincell{c}{$\mathcal{O}(I_2N_YG_rJ+I_3N_YJL(P-1))$} \\
		\cline{3-5}
		\hline
		\multirow{4}{*}{\tabincell{c}{Small-Timescale \\ Estimation}}	& \multirow{4}{*}{$\bar{Q}_1+\bar{Q}_2$}& \multirow{2}{*}{\tabincell{c}{Proposed On-Grid \\ Standard CS}} &\ding{172}&$\mathcal{O}(I_1\bar{Q}_1JG_r)$ 
		\\
		\cline{4-5}
		& & &\ding{173}&$\mathcal{O}(I_2\bar{Q}_1JG_r)$  \\
		\cline{3-5}
		& &\multirow{2}{*}{\tabincell{c}{Proposed Off-Grid \\ Standard CS}} &\ding{172}&$\mathcal{O}(\bar{Q}_1J(I_1G_r+I_3PL))$ \\
		\cline{4-5}
		& & &\ding{173}&$\mathcal{O}(\bar{Q}_1J(I_2G_r+I_3PL))$ \\
		\hline
		
	\end{tabular}
\end{table*}
  \subsubsection{Small-Timescale}
  The pilot overhead of this timescale is $\bar{Q}_1+\bar{Q}_2$. Generally, $\bar{Q}_1+\bar{Q}_2\ll Q_1+Q_2$ owing to the proposed MAE strategy. 
  Similar to the above analysis,
  recalling Eqn. (\ref{T2}), the per-iteration complexity of OMP is given by $\mathcal{O}(\bar{Q}_1JG_r+\bar{Q}_2JG_r)$. Since $\bar{Q}_1$ and $\bar{Q}_2$ are comparable, the complexity is simplified as $\mathcal{O}(\bar{Q}_1JG_r)$. Further, according to the previous analysis, the per-iteration complexity of off-grid refinement is $\mathcal{O}((\bar{Q}_1+\bar{Q}_2)PLJ)$. Hence, the total complexity of the off-grid OMP algorithm for small-timescale estimation is $\mathcal{O}(\bar{Q}_1J(I_1G_r+I_3PL))$. Along with this, the complexity of the off-grid EM-GAMP algorithm for small-timescale estimation is $\mathcal{O}(\bar{Q}_1J(I_2G_r+I_3PL))$.

\section{Conclusions}\label{S6}

This paper has effectively addressed the problem of 3D double-RIS-based separate channel estimation from two perspectives: 1) a low hardware cost and pilot overhead estimation protocol and 2) an efficient channel parameter recovery framework. For 1), we first propose a multi-user two-timescale channel estimation protocol based on the active RIS architecture with only one RF chain, where the multi-user uplink training strategy is proposed for performance improvement. 
Notably, our proposed two-timescale estimation protocol decreases the pilot overhead from two aspects.
Using the conventional estimation protocol, the demanding times for
estimating the large-timescale channels are equal to that for estimating the small-timescale channels. In contrast, the two-timescale estimation protocol requires less times to estimate the large-timescale channels, so as to decrease the pilot overhead.
 Moreover, simulations have demonstrated that our proposed MAE-based strategy can effectively reduce the pilot overhead for
 small-timescale channel estimation.
 For 2), the SVD-MMV-CS framework has been proposed for computing efficiently for the large-timescale estimation problem. Not only that, simulation results demonstrate that this framework has better NMSE performance than the conventional Kronecker framework and the SVD-CS framework. Furthermore, the LoS-aided M-EM-GAMP algorithm and its off-grid version are presented to estimate the channel parameters based on the SVD-MMV-CS framework. Simulation results have shown that the NMSE performance of EM-GAMP-based methods is better than that of OMP-based methods in different simulation scenarios.

\begin{appendices}
	
\section{  }\label{appendixA}
Recalling Eqn. (\ref{YPHI}), we know that $\mathcal{C}(\mathbf{Y})\subset\mathcal{C}(\bm{\Phi}_1\bm{\Delta})\subset\bm{\Phi}_1\mathcal{C}(\bm{\Delta})$, as well as $\mathcal{C}(\mathbf{Y}^H)\subset\mathcal{C}(\bm{\Phi}_2\bm{\Delta}^H)\subset\bm{\Phi}_2\mathcal{C}(\bm{\Delta}^H)$. Since $\mathbf{e}_{1,k}$ and $\mathbf{e}_{2,k}$ are the column vectors of the SVD of $\mathbf{Y}$ and $\mathbf{Y}^H$, respectively, such that $\mathbf{e}_{1,k}\in\mathcal{C}(\mathbf{Y})\subset\bm{\Phi}_1\mathcal{C}(\bm{\Delta})$ and $\mathbf{e}_{2,k}\in\mathcal{C}(\mathbf{Y}^H)\subset\bm{\Phi}_2\mathcal{C}(\bm{\Delta}^H)$. Therefore, the solution of $ {\rm min}\{ \Vert\bm{\delta}_{i,k} \Vert_0+ \Vert\mathbf{e}_{i,k}-\bm{\Phi}_i\bm{\delta}_{i,k}\Vert_2^2\}$, i.e., $\widehat{\bm{\delta}}_{i,k}$ is $P$-sparse due to $\bm{\Phi}_1\widehat{\bm{\delta}}_{1,k}\approx\mathbf{e}_{1,k}\in\bm{\Phi}\mathcal{C}(\bm{\Delta})$ and $\bm{\Phi}_2\widehat{\bm{\delta}}_{2,k}\approx\mathbf{e}_{2,k}\in\bm{\Phi}\mathcal{C}(\bm{\Delta}^H)$. In particular, the relation of $\forall k,\widehat{\bm{\delta}}_{1,k}/\widehat{\bm{\delta}}_{2,k} \in\mathcal{C}({\bm{\Delta}})/\mathcal{C}({\bm{\Delta}}^H)$ indicates that $\{\widehat{\bm{\delta}}_{1,k}\}_{k=1}^R/\{\widehat{\bm{\delta}}_{2,k}\}_{k=1}^R$ have the same row sparsity support w.r.t. $\bm{\Delta}/\bm{\Delta}^H$. Hence, the issue of Eqn. (\ref{E}) holds.
Considering the accurate estimation of $\bm{\Delta}^{\rm M}_i,i\in\{1,2\}$ such that $\widehat{\mathbf{Y}}=\mathbf{Y}=\widehat{\mathbf{E}}_1\widehat{\mathbf{E}}_2^H$, hence the following holds:
\begin{equation}
	\setlength{\abovedisplayskip}{3.2pt}
	\bm{\Phi}_1\widehat{\bm{\Delta}}^{\rm M}_1(\widehat{\bm{\Delta}}_2^{\rm M})^H\bm{\Phi}_2^H=\bm{\Phi}_1\bm{\Delta}\bm{\Phi}_2^H.
	\setlength{\belowdisplayskip}{3.2pt}
\end{equation}
However, the above equation cannot derive $\widehat{\bm{\Delta}}^{\rm M}_1(\widehat{\bm{\Delta}}_2^{\rm M})^H=\bm{\Delta}$ directly since $\bm{\Phi}_1/\bm{\Phi}_2$ may have not full column rank.
Next, an indirect proof for Eqn. (\ref{DD}) is provided. First, we assume that Eqn. (\ref{DD}) does not hold, i.e., $\bm{\varepsilon}=\bm{\Delta}-\widehat{\bm{\Delta}}\neq\mathbf{0}$. Let $\bm{\varepsilon}=\sum_{r^\prime=1}^{R^\prime}\bm{\mathcal{D}}_{1,r^\prime}\bm{\mathcal{D}}_{2,r^\prime}^H$ be a $R^\prime$-rank decomposition of $\bm{\varepsilon}$ such that $\bm{\mathcal{D}}_{1,r^\prime}\in\mathcal{C}(\bm{\Delta})$ and $\bm{\mathcal{D}}_{2,r^\prime}\in\mathcal{C}(\bm{\Delta}^H)$. This is because $\mathcal{C}(\bm{\varepsilon})\subset\mathcal{C}(\bm{\Delta})$ and $\mathcal{C}(\bm{\varepsilon}^H)\subset\mathcal{C}(\bm{\Delta}^H)$. Furthermore, we define $\bm{\mathcal{Y}}=\bm{\Phi}_1\bm{\varepsilon}\bm{\Phi}_2^H$ such that ${\rm rank}(\bm{\mathcal{Y}})\leq R^\prime$. Owing to the NSP, we have ${\rm dim}(\bm{\Phi}_1\mathcal{C}(\bm{\varepsilon}))={\rm dim}(\bm{\Phi}_2\mathcal{C}(\bm{\varepsilon}^H))=R^\prime={\rm rank}(\bm{\mathcal{Y}})$. Therefore, $\bm{\mathcal{Y}}$ has rank $R^\prime$ so that $\bm{\mathcal{Y}}\neq \mathbf{0}$. In fact, we know that $\bm{\mathcal{Y}}=\mathbf{Y}-\widehat{\mathbf{Y}}=0$. This is contrary to the conclusion under the assumption of $\bm{\varepsilon}\neq\mathbf{0}$, hence Eqn. (\ref{DD}) holds. The proof of \textbf{Theorem 1} is completed.

\section{  }\label{appendixB}
Here, we give the proof of \textbf{Theorem 2}. 
Since \begin{equation}
	\setlength{\abovedisplayskip}{3.2pt}
	\left\| \mathbf{Y}-\sum_{b=1}^Pw_b\mathbf{A}_b\right\|_F^2=\left\| {\rm vec}\left(\mathbf{Y}-\sum_{b=1}^Pw_b\mathbf{A}_b\right)\right\|_2^2,
	\setlength{\belowdisplayskip}{3.2pt}
\end{equation}
then the MLS issue is expressed as
$
	\underset{\mathbf{w}}{\rm arg \ min} \ \left\| {\rm vec}\left(\mathbf{Y}-\sum_{b=1}^Pw_b\mathbf{A}_b\right)\right\|_2^2
$.
This can be solved by the conventional LS algorithm with 
$
		{\rm vec}(\mathbf{Y})-\sum_{b=1}^Pw_b{\rm vec}(\mathbf{A}_b)={\rm vec}(\mathbf{Y})-\mathbf{G}\mathbf{w},
$
where $\mathbf{G}=[{\rm vec}(\mathbf{A}_1),{\rm vec}(\mathbf{A}_2),\cdots,{\rm vec}(\mathbf{A}_P)]\in\mathbb{C}^{MR\times P}$. Thus, we have
$
	\mathbf{w}=(\mathbf{G}^H\mathbf{G})^{-1}\mathbf{G}^H{\rm vec}(\mathbf{Y})
$.
$\textbf{Theorem 2}$ has been proved.
\end{appendices}
	
\vspace{-0.55em}

\bibliographystyle{IEEEtran}
\bibliography{reference.bib}

\vspace{12pt}

\end{document}